\documentclass[floatfix, nofootinbib]{emulateapj}
\usepackage{epsfig,subfigure,amsmath}
\usepackage{color}

\definecolor{purple}{rgb}{1,0,1}

\newcommand{\hmpc}{$h^{-1}$Mpc}

\bibpunct[; ]{(}{)}{;}{a}{}{,}

\begin{document}

\title{A first application of the Alcock-Paczynski test
       to stacked cosmic voids}
\author{ P.~M.~Sutter$^{1,2,3,4}$,
         Guilhem Lavaux$^{5,6}$,
         Benjamin~D.~Wandelt$^{2,3,1,7}$, and
         David~H.~Weinberg$^{4,8}$\\
 {~}\\
$^{1}$ Department of Physics, University of Illinois at Urbana-Champaign, Urbana, IL 61801\\
$^{2}$ UPMC Univ Paris 06, UMR7095, Institut d'Astrophysique de Paris, F-75014, Paris, France \\
$^{3}$ CNRS, UMR7095, Institut d'Astrophysique de Paris, F-75014, Paris, France
\\
$^{4}$ Center for Cosmology and Astro-Particle Physics, Ohio State University, Columbus, OH 43210\\
$^{5}$ Department of Physics \& Astronomy, University of Waterloo, Waterloo,
ON,  N2L 3G1 Canada \\
$^{6}$ Perimeter Institute for Theoretical Physics,
Waterloo, ON, N2L 2Y5, Canada \\
$^{7}$ Department of Astronomy, University of Illinois at Urbana-Champaign, Urbana, IL 61801\\
$^{8}$ Department of Astronomy, Ohio State University, Columbus, OH 43210\\
}

\begin{abstract}
We report on the first application of the Alcock-Paczynski test 
to stacked voids in spectroscopic galaxy redshift surveys.
We use voids from the Sutter et al. (2012) void catalog,
which was derived from
the 
Sloan Digital Sky Survey Data Release 7 main sample and luminous red galaxy
catalogs.
The construction of that void catalog removes potential 
shape measurement bias by using
a modified version of the {\tt ZOBOV} algorithm 
and by removing voids near survey boundaries and masks.
We apply the shape-fitting procedure presented in 
Lavaux \& Wandelt (2012) to ten void stacks out to redshift $z=0.36$.
Combining these measurements, we determine the mean cosmologically induced
``stretch'' of voids in three redshift bins, with $1\sigma$ errors
of 5-15\%.  The mean stretch is consistent with unity, providing no
indication of a distortion induced by peculiar velocities.
While the statistical errors are too large to detect the Alcock-Paczynski
effect over our limited redshift range, this proof-of-concept
analysis defines procedures that can be applied to larger spectroscopic
galaxy surveys at higher redshifts to constrain dark energy
using the expected statistical isotropy of structures that are
minimally affected by uncertainties in galaxy velocity bias.
\end{abstract}


\maketitle

\section{Introduction}
\label{sec:introduction}

Characterizing the nature and history of dark energy is
perhaps the greatest challenge in the near future of observational
cosmology.
Many elementary probes now strive to
distinguish a cosmological constant from
alternative theories of dynamical dark energy or modified gravity.
Most probes rely on
``standard candles,'' such as Type Ia
supernovae~\citep[e.g.,][]{AlderingGreg2002}, or
``standard rulers,'' such as radio galaxy 
diameters~\citep{Daly2009} or baryon
acoustic oscillations (BAO)~\citep[e.g.,][]{Eisenstein2005, Blake2011, Beutler2011, Anderson2012, MehtaKushalT.2012}.
Reviews of dark energy probes, current constraints and
 forecasts for future experiments include
~\citet{LinderEricV.2003},
~\citet{AlbrechtAndreas2006},
~\citet{Frieman2008},
and~\citet{Weinberg2012}.

Over 30 years ago Alcock \& Paczynski (1979, hereafter AP)
proposed an elegant alternative approach based on a hypothetical
population of idealized spheres.
Their key insight was that
since galaxy
spatial positions are inferred from both their angular positions \emph{and}
redshifts, these spheres will appear anisotropic if one adopts
an incorrect spacetime metric.
Specifically, because line-of-sight
distances scale with the inverse Hubble parameter $H^{-1}(z)$ and
transverse distances scale with the angular diameter distance $D_A(z)$,
their ratio, or \emph{stretch}, measures the value of the product $H(z)D_A(z)$.
In practice, the AP test requires only \emph{statistical} isotropy of the
observed structures, so the test can be implemented with measures
of quasar, galaxy, or Ly$\alpha$ forest clustering, or features 
in the redshifted 21~cm spectrum~\citep[e.g.,][]{Hui1999, McDonald1999, 
EriksenK.A.2005, Nusser2005, 
Kim2007a, Blake2011, Reid2012}.
In this paper we apply the AP
test to the~\citet{Sutter2012} catalog of voids in the galaxy
redshift surveys of the Sloan Digital Sky Survey
(SDSS;~\citealt{York2000}).

To date, most applications of the AP test have focused on
the autocorrelation function or power spectrum
~\citep[e.g.,][]{Ballinger1996,Matsubara1996,Matsubara2004}.
Specifically,
the clearest detections of the AP effect have been found in
the two-point correlations of galaxies in the
WiggleZ survey~\citep{Blake2011} and the Baryon Oscillation
Spectroscopic Survey~\citep{Reid2012}.
A successful application of the AP test requires handling
the large systematic uncertainties caused by peculiar motions,
which introduce redshift-space
anisotropy that must be disentangled from the AP effect itself.
Uncertainties in the peculiar velocity corrections limit the
~\citet{Blake2011} and~\citet{Reid2012} studies to large, quasi-linear scales,
where the statistical uncertainties are relatively large.
An attempt was made recently to apply the AP test to close 
galaxy pairs~\citep{Marinoni2010}, but as~\citet{Bueno2012} 
point out this method has serious shortcoming due to 
dynamics at small scales. Additionally, the 
analysis of~\citet{Jennings2012} indicates that this method provides 
relatively weak constraints.

Cosmic voids provide an attractive alternative for applying
the AP test, as first proposed by~\citet{Ryden1995} and discussed
extensively by~\citet{LavauxGuilhem2011}.
Voids are the large, underdense regions that occupy a large fraction of the
volume of the Universe and are a natural consequence of the hierarchical
growth of structure~\citep{Hausman1983,Thompson2011}.
While peculiar velocities modestly affect void
shapes~\citep{Ryden1996,Maeda2011,LavauxGuilhem2011}, voids avoid the
regions of high velocity dispersion that have such a large impact
on the redshift-space correlation function and power spectrum.
Indeed, modeling of peculiar velocities in voids is particularly
straightforward since they are still in the quasi-linear regime.
In addition, the scale of voids is fairly small, with typical
comoving radii $\sim 10$~\hmpc~in a densely sampled survey, and
they have a large filling factor (i.e., they occupy a majority of the volume
of the Universe), amplifying their statistical power
relative to other techniques.
We can therefore measure the
mean void shape with high precision in a large volume survey.
~\citet{LavauxGuilhem2011} showed that a statistics-limited
void AP test can dramatically improve the dark energy constraints
from the redshift survey planned for the Euclid
satellite~\citep{Laureijs2011};
the AP test outperforms the BAO
constraints from the same survey, even though BAO constraints leverage 
a known standard ruler, because the scale of the voids
is so much smaller than the BAO scale, yielding correspondingly
more precise measurements.

We take the void sample for this analysis from the catalog
described in~\citep{Sutter2012}.
That work constructed a void catalog from the main galaxy redshift survey
~\citep{Strauss2002} and the luminous red galaxy (LRG) redshift survey
\citep{Eisenstein2001} of the SDSS Seventh Data Release
(DR7;~\citealt{Abazajian2009}). We identify voids
using a modified version of
the Voronoi-based {\tt ZOBOV} algorithm~\citep{Neyrinck2008}.
To compensate for the significant Poisson sampling noise in shape
measurements of individual voids~\citep{Shoji},
we instead measure the mean void shape by ``stacking''
the galaxy distributions of our identified voids in bins of
redshift and radius.
The SDSS LRG survey is sparse, so at $z \ga 0.2$ we can only
identify large voids, which are limited in number.
The combination of moderate redshift leverage and limited statistics
prevents us from making a secure detection of the AP effect in
this sample, but our proof-of-concept analysis addresses
many of the practical issues that will also arise in future
data sets at higher redshift.

In the following section
we give a brief overview of our method for measuring void shapes and
applying the AP test. We review the properties of
the~\citet{Sutter2012} void catalog in Section~\ref{sec:catalogs}, followed
by a presentation of the stacked voids in Section~\ref{sec:stacks}.
We estimate the uncertainty in the stretch measurement
and present the AP test as applied to our void stacks in
Section~\ref{sec:ellipticity}.
Finally, we offer concluding remarks and a brief discussion of
prospects for future surveys in Section~\ref{sec:conclusions}.

\section{Measuring void stretch \& the AP test}
\label{sec:aptest}

Our definitions and procedures closely follow those described
by~\citet{LavauxGuilhem2011},
 who present tests on N-body simulations and forecasts
for surveys such as BOSS~\citep{Dawson2012} and Euclid~\citep{Laureijs2011}.
Given a galaxy's sky latitude $\theta$, sky longitude $\phi$, and 
redshift $z$, we transform to a hybrid coordinate system
\begin{eqnarray}
  x' & = & \frac{cz}{H_0} \cos{\phi} \cos{\theta}, \nonumber \\
  y' & = & \frac{cz}{H_0} \sin{\phi} \cos{\theta},  \\
  z' & = & \frac{cz}{H_0} \sin{\theta}, \nonumber
\label{eq:transform}
\end{eqnarray}
where $c$ is the speed of light and $H_0$ is the Hubble parameter 
at redshift $z=0$.
Note that our coordinate transformation preserves relative distances; 
in effect, we are performing a slightly modified version of the AP test
where we measure shapes directly in redshift space (see~\citealt{Ryden1995} 
for a discussion).
Since the AP test only applies to two dimensions (the extent along the 
line of sight and the projected angular extent), we may project 
positions within the void onto a plane:
\begin{eqnarray}
  d_v & = & \sqrt{x_{\rm rel}^2 + y_{\rm rel}^2} \\
  z_v & = & |z_{\rm rel}| \nonumber,
\label{eq:transformation}
\end{eqnarray}
where $(x_{\rm rel}, y_{\rm rel}, z_{\rm rel})$ are the galaxy coordinates 
relative to the void barycenter ${\bf X}_v$:
\begin{equation}
  {\bf x}_{\rm rel} \equiv {\bf x}' - {\bf X}_v.
\end{equation}

When stacking voids we place all void barycenters at a common 
point and 
rotate the galaxies within each void about a specified axis
 so that they all share a common line of sight. 
We then pixelize the density using 10 bins 
within the maximum void size in that stack, 
which helps smooth spurious density fluctuations.

For each stacked void, we assume an inner radial profile with form
\begin{equation}
  \frac{n(r)}{\bar{n}} = A_0 + A_3 \left( \frac{r}{R_v} \right)^4,
\label{eq:radial}
\end{equation}
where $r = |{\bf x}_{\rm rel}|$, $R_v$ is the void radius (which for our 
void definition is the effective radius, or the radius of the sphere 
which has the same volume as the Voronoi-based void volume), 
$\bar{n}$ is mean galaxy number density within the given sample, 
and $A_0$ and $A_3$ are free parameters.
As discussed by~\citep{Sutter2012}, due to galaxy bias and Poisson fluctuations 
caused by the sparseness with which 
galaxies sample the underlying density distribution 
the profiles seen in observations 
are steeper than those in dark matter simulations, so we use a steeper 
form for a fitted curve.
Using this radial profile we fit to an ellipse given by
\begin{equation}
  n(d_v,z_v) = {\rm min} 
    \left(n_0 + \left((d_v/a_d)^2 + (z_v/a_z)^2 \right)^{2}, n_{\rm max} \right),
\label{eq:ellipse}
\end{equation}
where $n_0$ is the density at the center of the stack, $a_d$ and 
$a_z$ are the semi-axes along the angular direction and redshift 
direction, respectively, and $n_{\rm max}$ is a maximum density 
value. 
Our fitting procedure requires an estimate of the uncertainty 
on a per-pixel basis.
Our bin smoothing described above allows us to assume 
that the fluctuations in each pixel 
are Gaussian with minimal covariance between pixels. 
We assume with two separate standard deviations
depending on the location inside the stack:
\begin{equation}
  \sigma(d,z) = \begin{cases} 
    \sigma_0 \sqrt{\frac{1 h^{-1} {\rm Mpc}}{d_v}}, & 
       \mbox{if } n(d_v,z_v) < n_{\rm max} \\ 
     \sigma_1, & 
       \mbox{otherwise}
         \end{cases}
\label{eq:stdev}
\end{equation}
This gives us a per-pixel sampling uncertainty within the void radius 
and a fixed standard deviation outside the void. We keep the latter 
uncertainty fixed because the regions outside the void carry large statistical 
weight but are largely unimportant to the fit.
The factor of $1/d$ accounts for the cylindrical averaging of pixels 
as we form the stack.
The values of $\sigma_0$ and $\sigma_1$ may be different among different 
samples, since these give a measure of the relative level of 
Poisson fluctuations. 
However, within a sample we expect --- and find --- that these values 
are consistent across multiple stacks. 

We truncate our stack at $R_{\rm cut} \equiv 3R_v$. 
We run a Monte Carlo Markov chain to explore the four parameters 
of Equation~(\ref{eq:ellipse}), the two standard deviations of 
Equation~(\ref{eq:stdev}), and their uncertainties.  
The likelihood that we must then explore takes the form
\begin{equation}
  \chi^2 = \sum_{i=1}^{N_d} \sum_{j=1}^{N_z} \left( 
   \frac{(n(d_i,z_i) - n_{i,j})^2}
   {\sigma^2(d_i,z_i)} 
   + 2 \log{\sigma(d_i,z_i)},
   \right)
\label{eq:likelihood}
\end{equation}
where $N_d$ is the number of pixels in the angular direction, 
and $N_z$ is the number of pixels in the redshift direction.
The values $d_i$ and $z_i$ are the values of the coordinates $d_v$ and $z_v$ 
at their respective indices. The exploration of this likelihood 
gives us both the measurement of the void ellipticity and 
the overall uncertainty associated with each stack.
The analysis of~\citet{LavauxGuilhem2011} found that the error bars 
produced from this method were consistent with the level of scatter 
among independent N-body simulations, though we will conclude
below that they underestimate the errors in our data set.

We translate these ellipticities into a cosmological measurement 
by applying the AP test, in which we measure the ratio of the length 
along the line of sight to the angular diameter of each stacked void.
We will call this ratio the void \emph{stretch}.
What follows is a brief discussion of the stretch as a function of redshift; 
see~\citet{LavauxGuilhem2011} for a more complete derivation.

We wish to take the ratio of a void length along the line of sight $\delta z_v$ 
to its projected angular extent $\delta d_v$.
In the simple coordinate system of Equation~(\ref{eq:transform}),
the projected angular extent is related to the angular extent by
$\delta d_v \equiv c z \delta \theta /H_0$. The angular extent in turn 
depends on cosmology via the angular diameter distance $D_A(z)$:
\begin{equation}
  \delta \theta = \frac{\delta r_v}{D_A(z)},
\end{equation}
where $\delta r_v$ is the comoving radial extent of the void.
 In a flat universe, the angular 
diameter distance is equal to the comoving line of sight distance $D_c(z)$:
\begin{equation}
  D_A(z) = D_c(z) = \frac{c}{H_0} \int_0^z \frac{d z'}{E(z')},
\label{eq:angdiam}
\end{equation}
where $E(z) \equiv H(z)/H_0$.
Combining these gives the expression
\begin{equation}
  \delta d = \frac{c}{H_0} \frac{z \delta r_v}{D_A(z)}.
  \label{eq:deld}
\end{equation}

The comoving line of sight distance $\delta z_v$ is also related to 
$D_c(z)$, and hence $D_A(z)$ 
in a flat universe, via
\begin{equation}
  \delta z_v = \frac{\delta l_v}{d D_A/dz},
\end{equation}
where $\delta l_v$ is the comoving distance along the line of sight. 
Taking the derivative allows us to identify 
\begin{equation}
  \delta z_v = \frac{H_0}{c} E(z) \delta l_v.
  \label{eq:delz}
\end{equation}

In an isotropic universe a stacked void should have the same 
extent in all directions; thus, its angular extent should equal its 
comoving distance along the line of sight. 
This allows us to assume $\delta l_v = \delta r_v$.
Combining Equations~(\ref{eq:deld}) and~(\ref{eq:delz}) above leads to our desired 
ratio:
\begin{equation}
  \frac{\delta z_v}{\delta d_v} = \left( \frac{H_0}{c} \right)^2 
                               \frac{D_A(z) E(z)}{z}.
  \label{eq:ap}
\end{equation}
We identify the void stretch, denoted by $e_v(z)$ for a void at redshift 
$z$, as
\begin{equation}
  e_v(z) \equiv \frac{c}{H_0} \frac{\delta z_v}{\delta d_v}.
  \label{eq:stretch}
\end{equation}

We measure this stretch by taking the fitted ellipse parameters of 
Equation~(\ref{eq:ellipse}) and identifying $\delta z_v$ as $a_z$ and 
$\delta d_v$ as $a_d$.
As we stack voids within redshift bins, we assume that the stack provides 
a measurement of the \emph{average} stretch in that bin, 
$\left< \delta z_v / \delta d_v \right>$, and we will compare that to the 
average expected stretching in that bin weighted by the void distribution:
\begin{equation}
  \overline{e_v}(z) = \frac{1}{N_v} \int_{z_i}^{z_i+\Delta z} {e_v(z')} N_v(z') dz',
\label{eq:avestretch}
\end{equation}
where the given bin runs from redshift $z_i$ to $\Delta z$, $N_v(z')$ is the 
number of voids in a given redshift slice, and $N_v$ is the total number of 
voids in the bin.

Throughout we will assume a 
flat universe with a cosmological constant,
which gives a Hubble equation of
\begin{equation}
  E(z,w_0,w_a) = \left( 
                 \Omega_m (1+z)^3 + \Omega_\Lambda
                 \right)^{1/2},
\label{eq:hubble}
\end{equation}
where
$\Omega_m$ and $\Omega_\Lambda$ are, respectively, the present-day matter and dark 
energy densities relative to the critical density.

\section{Void catalogs}
\label{sec:catalogs}

We take our void catalog from~\citet{Sutter2012}, which is based on 
volume-limited samples of the New York University Value-Added 
Galaxy Catalog~\citep{Blanton2005}. This catalog cross-matches galaxies
from SDSS~\citep{Abazajian2009} with other surveys using
improved photometric calibrations~\citep{Padmanabhan2008}.
We also use the 
LRG catalog of~\citet{Kazin2010}.
Table~\ref{tab:samples} summarizes the 
volume-limited samples used in this work.
Additionally, to improve our statistics by using as many voids as possible, 
we merge the four samples within $z<0.2$ into two 
samples: \emph{dim1+dim2} and \emph{bright1+bright2}.
For the smallest and largest voids in each sample these 
combinations violate our assumption that the voids are evenly 
distributed throughout each redshift bin. However, we do not use the 
very largest voids in any case, and the smallest voids have 
non-uniform redshift distributions within each sample, 
so we find that this does not strongly affect our results.

\begin{table*}
\centering
\caption{Data samples used in this work.}
\footnotesize
\tabcolsep=0.11cm
\begin{tabular}{ccccccc}
  Sample & Catalog & $M_{r,{\rm min}}$ & $z_{\rm min}$ & $z_{\rm max}$ & Number of Galaxies & Mean Spacing (\hmpc) \\
  \hline
  \hline
  dim1 & NYU VAGC & -18.9 & 0.0 & 0.05 & 63639 &  3 \\ 
dim2 & NYU VAGC & -20.4 & 0.05 & 0.1 & 156266 &  5 \\ 
bright1 & NYU VAGC & -21.35 & 0.1 & 0.15 & 113713 &  8 \\ 
bright2 & NYU VAGC & -22.05 & 0.15 & 0.2 & 43340 & 13 \\ 
lrgdim & LRGs & -21.2 & 0.16 & 0.36 & 67567 & 24 \\ 
lrgbright & LRGs & -21.8 & 0.36 & 0.44 & 15212 & 38 \\ 

\hline
\end{tabular}
\label{tab:samples}
\end{table*}

\citet{Sutter2012} produced void catalogs using a modified version of the 
void finder {\tt ZOBOV}~\citep{Neyrinck2008}, which maps the density 
using Voronoi tessellations~\citep{VandeWeygaert2007}
and collects these Voronoi cells into zones and voids using a watershed 
technique~\citep{Platen2007, Aragon-Calvo2010}.
This approach naturally identifies a full hierarchy of voids and sub-voids, 
which we will exploit to capture as many voids as possible. Additionally, 
the algorithm prevents voids from overlapping. 
To remove any potential shape measurement bias due to 
the presence of the mask, we choose 
the ``central'' catalog of voids, which are selected such that they 
could not possibly intersect any boundary or mask in the survey 
for any given rotation about their barycenters.

\section{Void stacks}
\label{sec:stacks}

We have many constraints for grouping voids into redshift and 
radius bins.
We choose redshift bins corresponding to the limits of the \emph{dim1+dim2}, 
\emph{bright1+bright2}, \emph{lrgdim}, and \emph{lrgbright} samples. 
Within each redshift bin, we divide the voids into bins of radius with the 
following objectives:
\begin{enumerate}
  \item \emph{Sufficient numbers -} Within each stack we require enough 
                                    voids to sufficiently smooth the projected
                                    density and increase the signal-to-noise 
                                    so that we can make reliable measurements.
                                    While this number is not fixed, we have 
                                    found empirically that we 
                                    require at least $\sim$15
                                    voids per stack for the subsequent 
                                    stretch measurement to converge reliably.
  \item \emph{Multiple radial bins -} As many independent measurements
                                  within the same redshift bin as possible
                                  allows us to account for scatter that 
                                  can develop in individual measurements by 
                                  taking uncertainty-weighted
                                  averages of these multiple
                                  measurements.
  \item \emph{Narrowness -} Each stack should have a narrow radial width; 
                            otherwise, the density profile will smooth out to 
                            the point that our shape fitting routine cannot 
                            reliably measure the ellipticity. Also, smaller 
                            voids tend to have more Poisson noise and can 
                            severely degrade the measurement when 
                            combined with larger voids.
  \item \emph{Even distribution -} The voids in each stack should be evenly 
                                   distributed in redshift so that we 
                                   reliably measure
                                   the mean ellipticity without bias.
\end{enumerate}

Given these conditions, we select four stacks from the \emph{dim1+dim2} and 
\emph{bright1+bright2} samples and two from the \emph{lrgdim} sample. We 
discard the \emph{lrgbright} sample because there are not enough voids to 
construct reliable stacks. 
The
void radius is used to assign voids to stacks. While the tessellation 
procedure gives the \emph{exact} volume for the given sampling, we may be
miscalculating the volume due to the sparseness of the sampling itself. 
However, we will choose sufficiently broad radial stacks such that this is not
a concern.
For each sample, our first stack begins at the 
mean galaxy separation $\bar{d} \equiv \bar{n}_d^{-1/3}$, 
which we take as the smallest resolvable 
void~\citep{Tikhonov2006, Platen2011}. 
Below 20~\hmpc~our bins have width 4~\hmpc; above this 
we switch to 8~\hmpc~widths to collect sufficient numbers of voids 
(the exception to this rule is the smallest stack of the 
\emph{bright1+bright2} sample, where we extend the width to 
include enough voids). 
We treat the \emph{lrgdim} sample slightly differently due to its 
much poorer resolution, small number of voids, and wide range of 
void sizes. For this sample we reject the smallest voids and 
construct one stack with
width 16~\hmpc~ and one with width 36~\hmpc. 
We do not 
include the very largest voids in each sample because they are difficult to 
reliably combine with smaller voids. However, our bins contain 
over $95\%$ of the voids in each sample, meaning that we are taking 
almost full advantage of the void information available in each 
catalog.

Within each stack we rescale the voids to the maximum void size in that 
stack; i.e, we multiply all positions relative to the void center by 
$R_v/R_{v, {\rm max}}$. This reduces the effects of Poisson scatter within the inner 
wall of the stacked void, improving our shape estimation.
Where we have combined voids from different samples, we normalize the profile 
of each void to the mean number density of galaxies in the sample before 
adding it to the stack.

\begin{figure*}[ht]
  \centering 
  {\includegraphics[type=pdf,ext=.pdf,read=.pdf,width=0.48\textwidth]{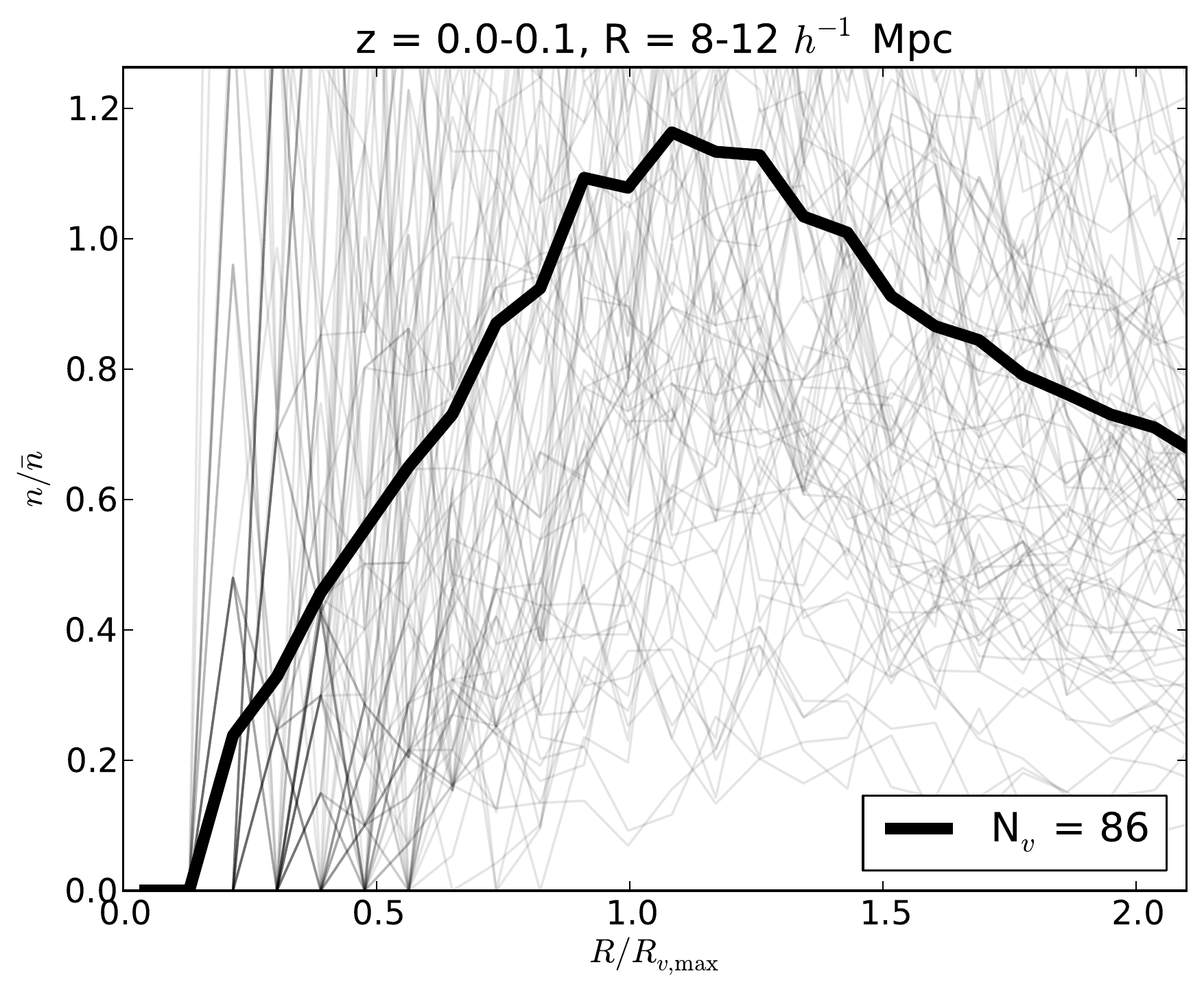}}
  {\includegraphics[type=pdf,ext=.pdf,read=.pdf,width=0.48\textwidth]{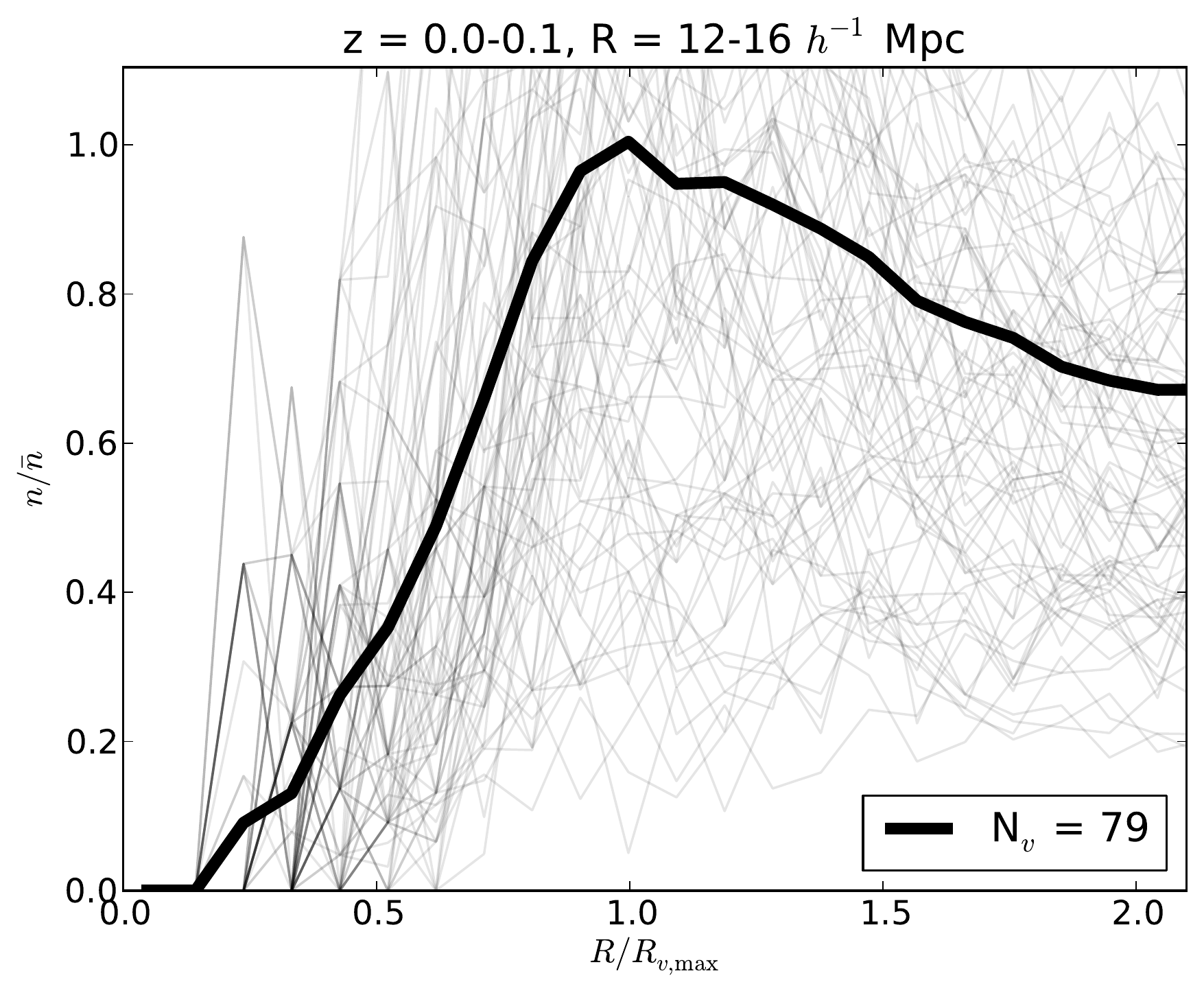}}
  {\includegraphics[type=pdf,ext=.pdf,read=.pdf,width=0.48\textwidth]{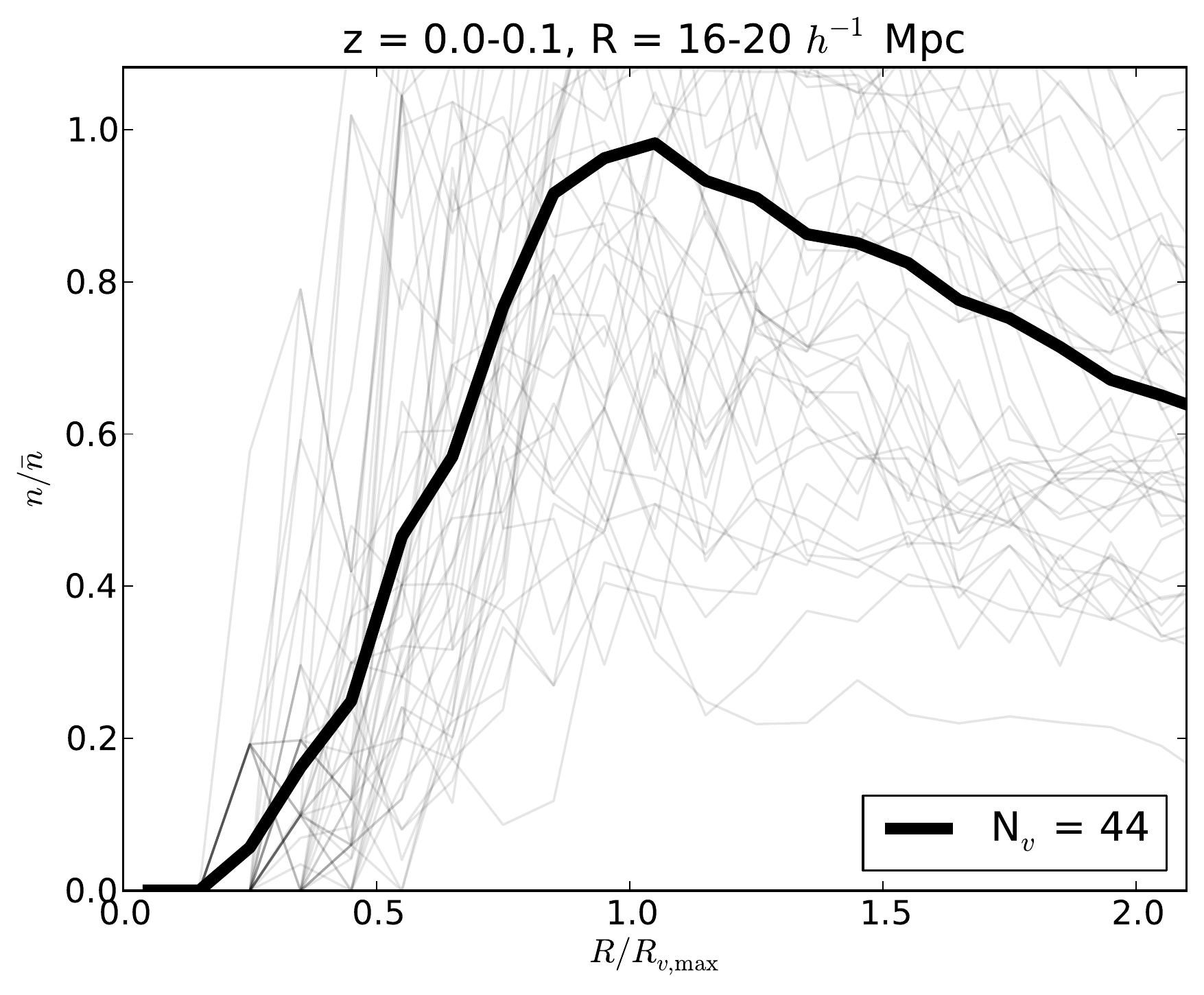}}
  {\includegraphics[type=pdf,ext=.pdf,read=.pdf,width=0.48\textwidth]{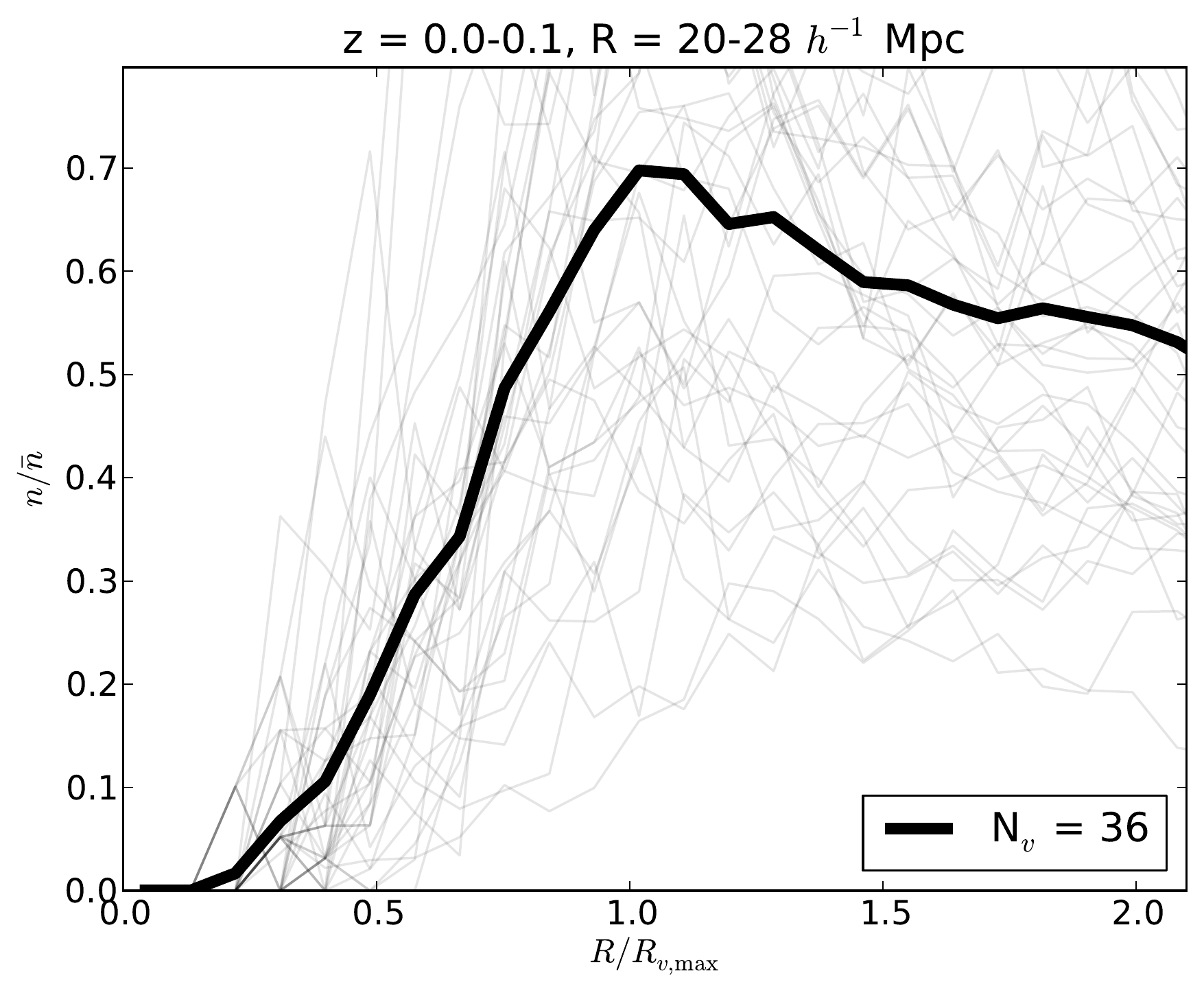}}
  \caption{\emph{Radial profiles at $0<z<0.1$ for the \emph{dim+dim2} sample.} 
           We show profiles of individual voids in gray and the profile of the 
           stacked void in black. The legend gives the number of voids in 
           each stack. We rescale each void to the maximum void size 
           within each stack.}
\label{fig:profile1ddim}
\end{figure*}

\begin{figure*} 
  \centering 
  {\includegraphics[type=pdf,ext=.pdf,read=.pdf,width=0.48\textwidth]{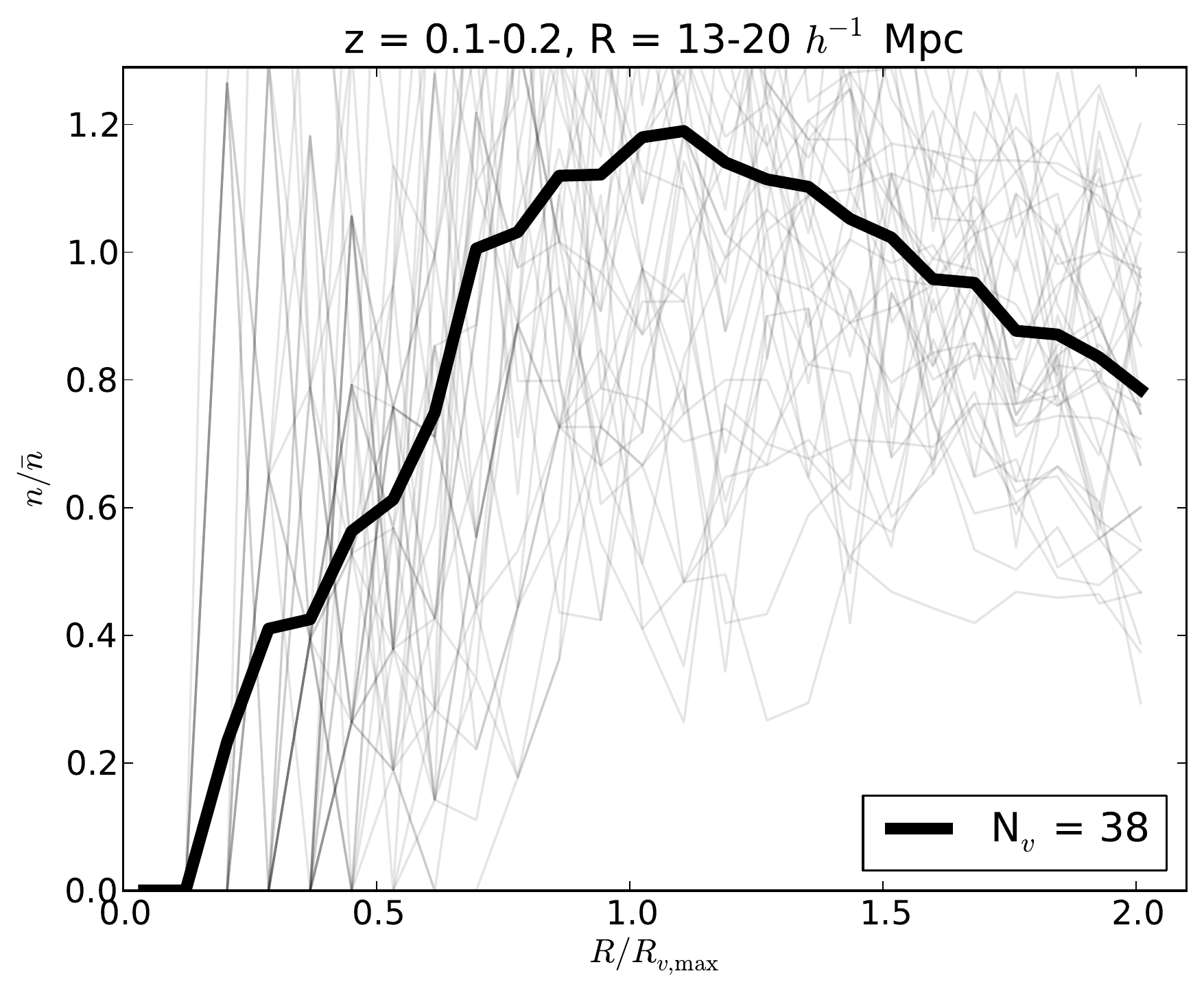}}
  {\includegraphics[type=pdf,ext=.pdf,read=.pdf,width=0.48\textwidth]{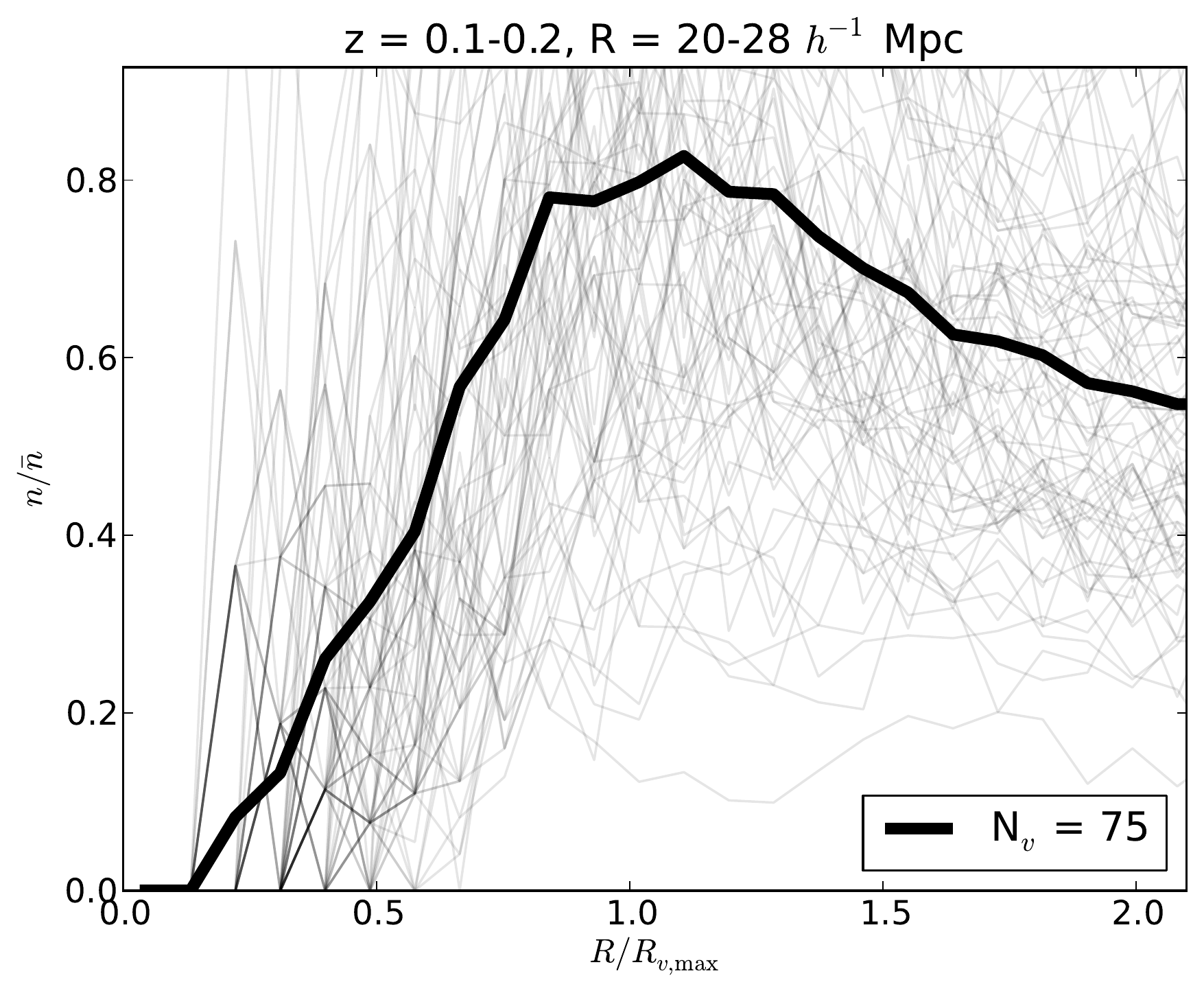}}
  {\includegraphics[type=pdf,ext=.pdf,read=.pdf,width=0.48\textwidth]{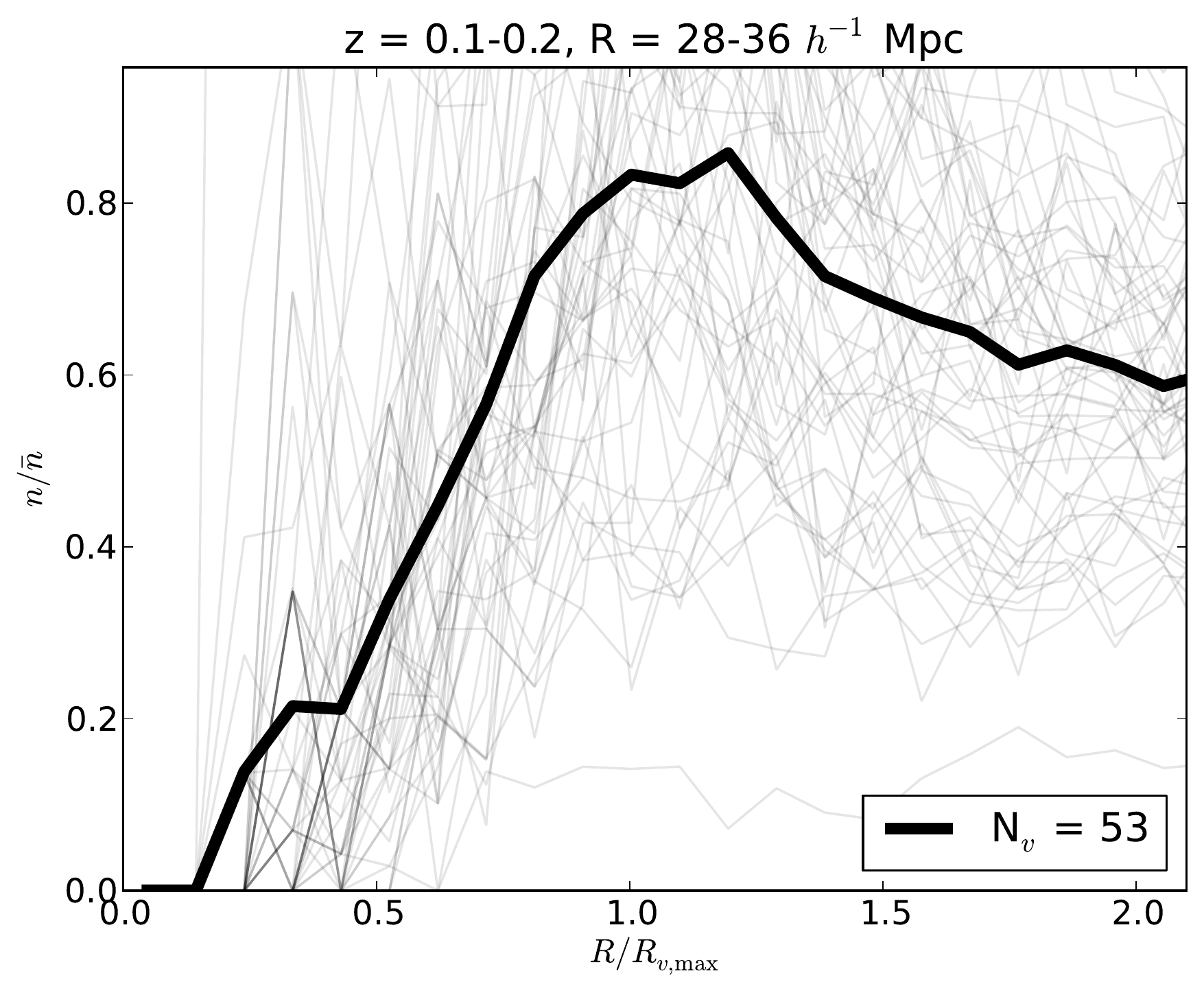}}
  {\includegraphics[type=pdf,ext=.pdf,read=.pdf,width=0.48\textwidth]{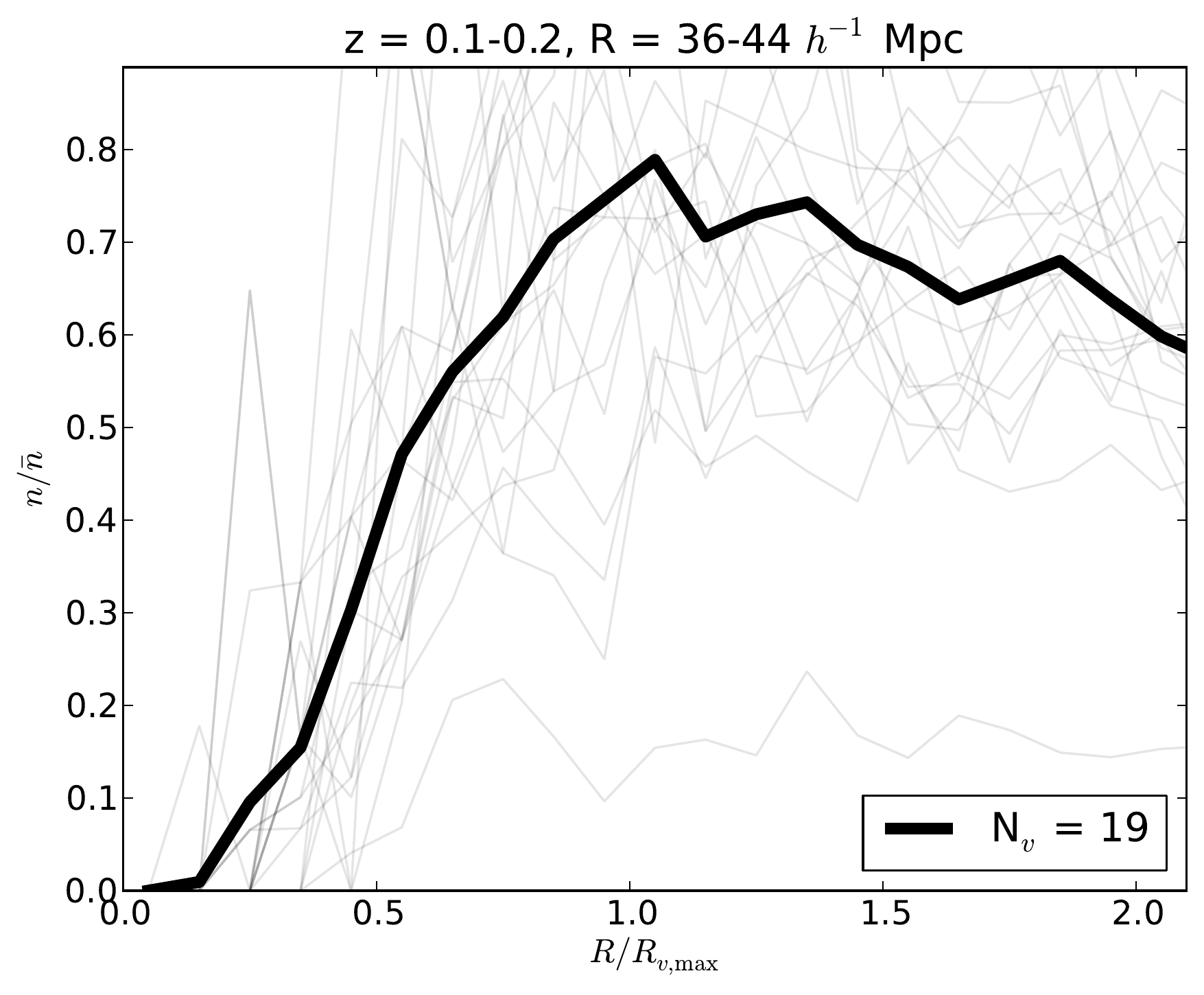}}
  \caption{\emph{Radial profiles at $0.1<z<0.2$ for the 
                 \emph{bright1+bright2} sample.} 
           See the caption for Figure~\ref{fig:profile1ddim} for a plot
           description.
           }
\label{fig:profile1dbright}
\end{figure*}

\begin{figure*} 
  \centering 
  {\includegraphics[type=pdf,ext=.pdf,read=.pdf,width=0.48\textwidth]{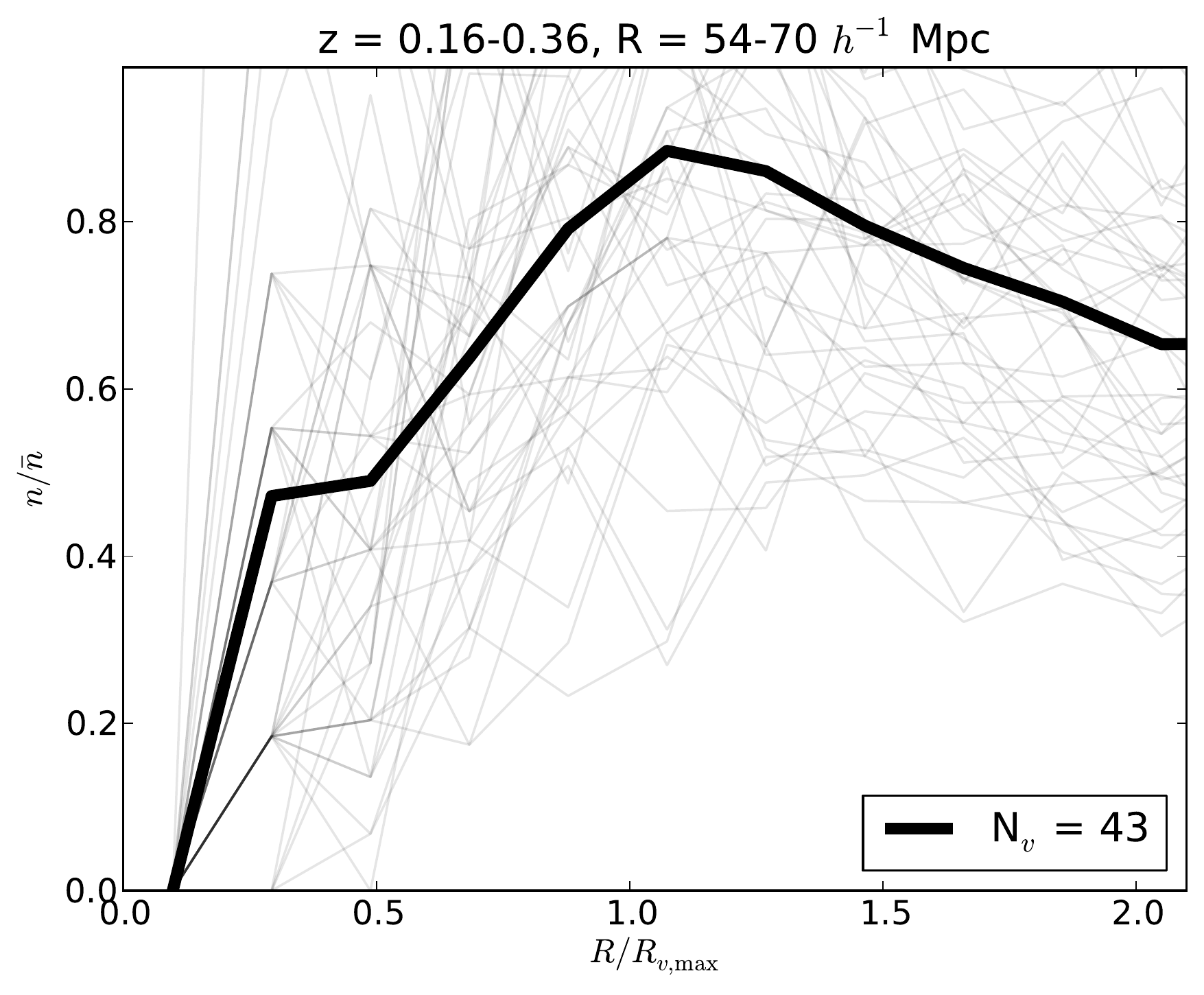}}
  {\includegraphics[type=pdf,ext=.pdf,read=.pdf,width=0.48\textwidth]{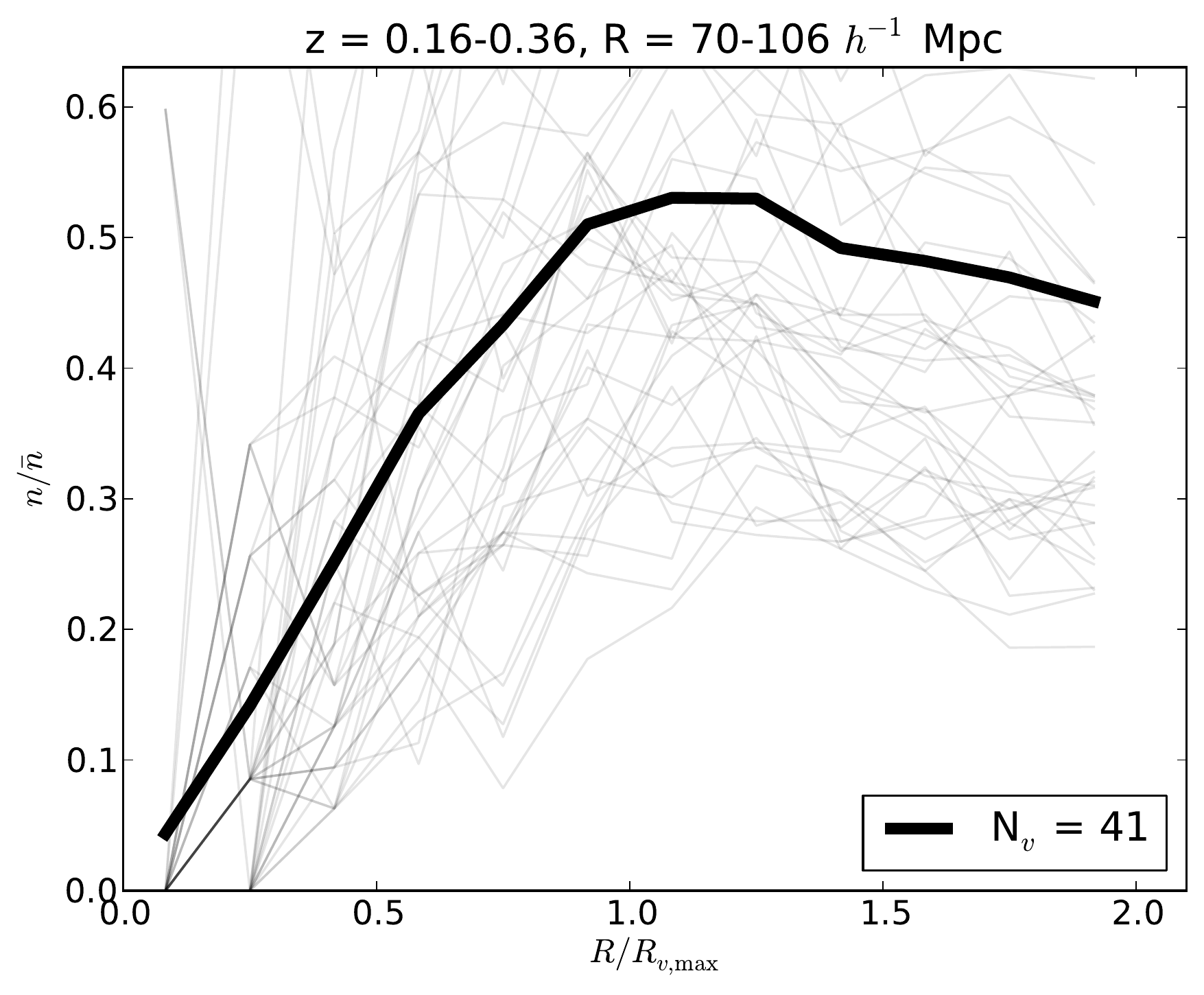}}
  \caption{\emph{Radial profiles at $0.16<z<0.36$ for the \emph{lrgdim} sample.}
           See the caption for Figure~\ref{fig:profile1ddim} for a plot
           description.
           }
\label{fig:profile1dlrgdim}
\end{figure*}

In Figures~\ref{fig:profile1ddim}, ~\ref{fig:profile1dbright}, 
and~\ref{fig:profile1dlrgdim} we show the one-dimensional 
radial profiles for each stack in the \emph{dim1+dim2}, \emph{bright1+bright2}, 
and \emph{lrgdim} samples, respectively. We construct these profiles by 
measuring the density within thin spherical shells. We also show the 
radial profiles of individual voids. Each individual void scatters about the
mean density of its sample at large radii.
It is only the stacked void which asymptotes to the
mean (as seen in~\citealt{Sutter2012}). By re-scaling voids, we move galaxies
outside the stack maximum radius, lowering the mean of the stack.

We immediately note the steep profiles relative to results from 
dark matter-only simulations. This is expected due to the effects of 
Poisson sampling and the biasing of galaxies as tracers of 
density and justifies our choice of a quartic radial profile for 
shape fitting.
These plots also highlight the necessity of stacking: attempts to 
measure the ellipticity of individual voids would be nearly impossible.
As in the analysis of~\citet{Sutter2012}, we see that the stacked 
voids greatly enhance the signal-to-noise and generate qualitatively 
similar profiles across many void sizes and redshift ranges.

\begin{figure*} 
  \centering 
  {\includegraphics[type=pdf,ext=.pdf,read=.pdf,width=0.48\textwidth]{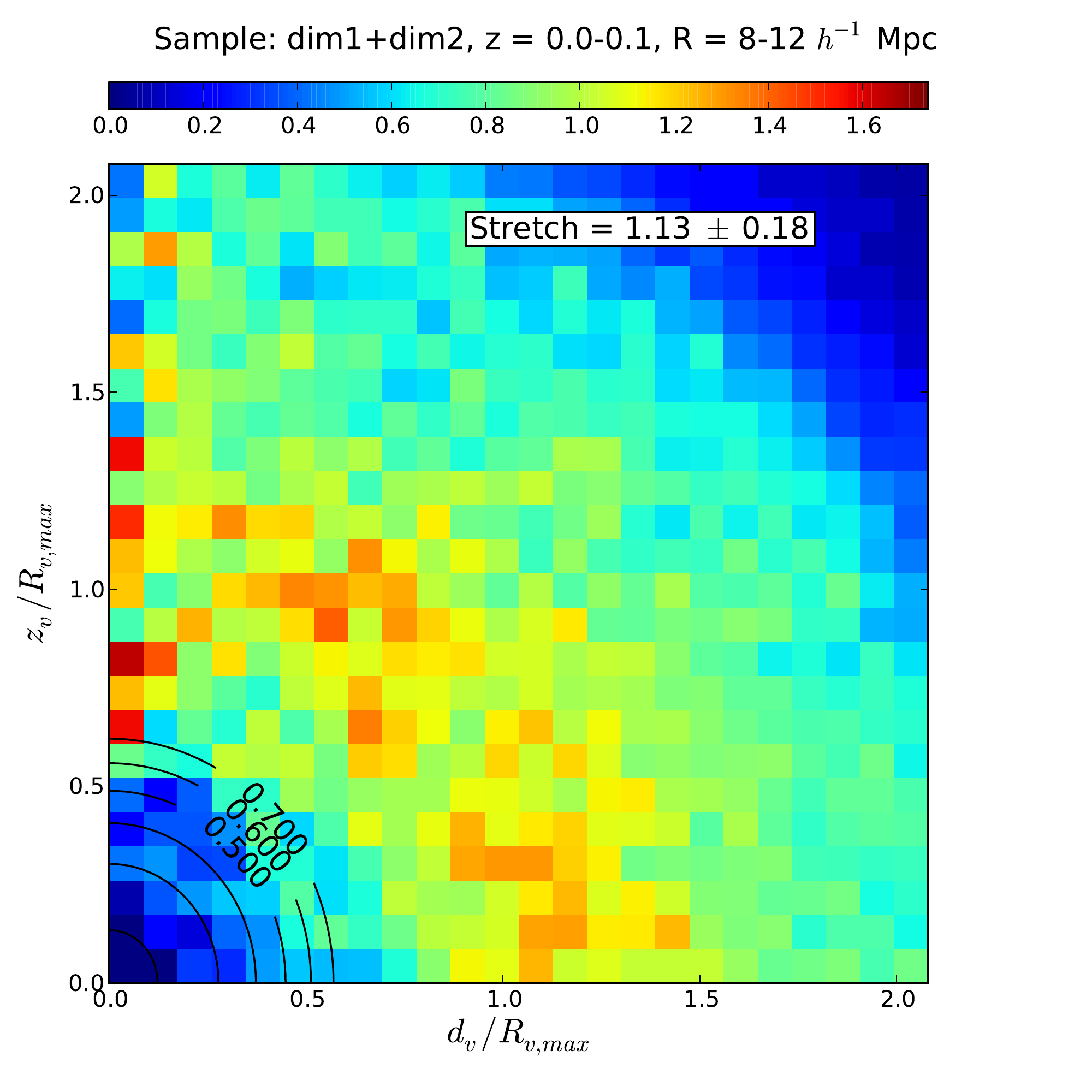}}
  {\includegraphics[type=pdf,ext=.pdf,read=.pdf,width=0.48\textwidth]{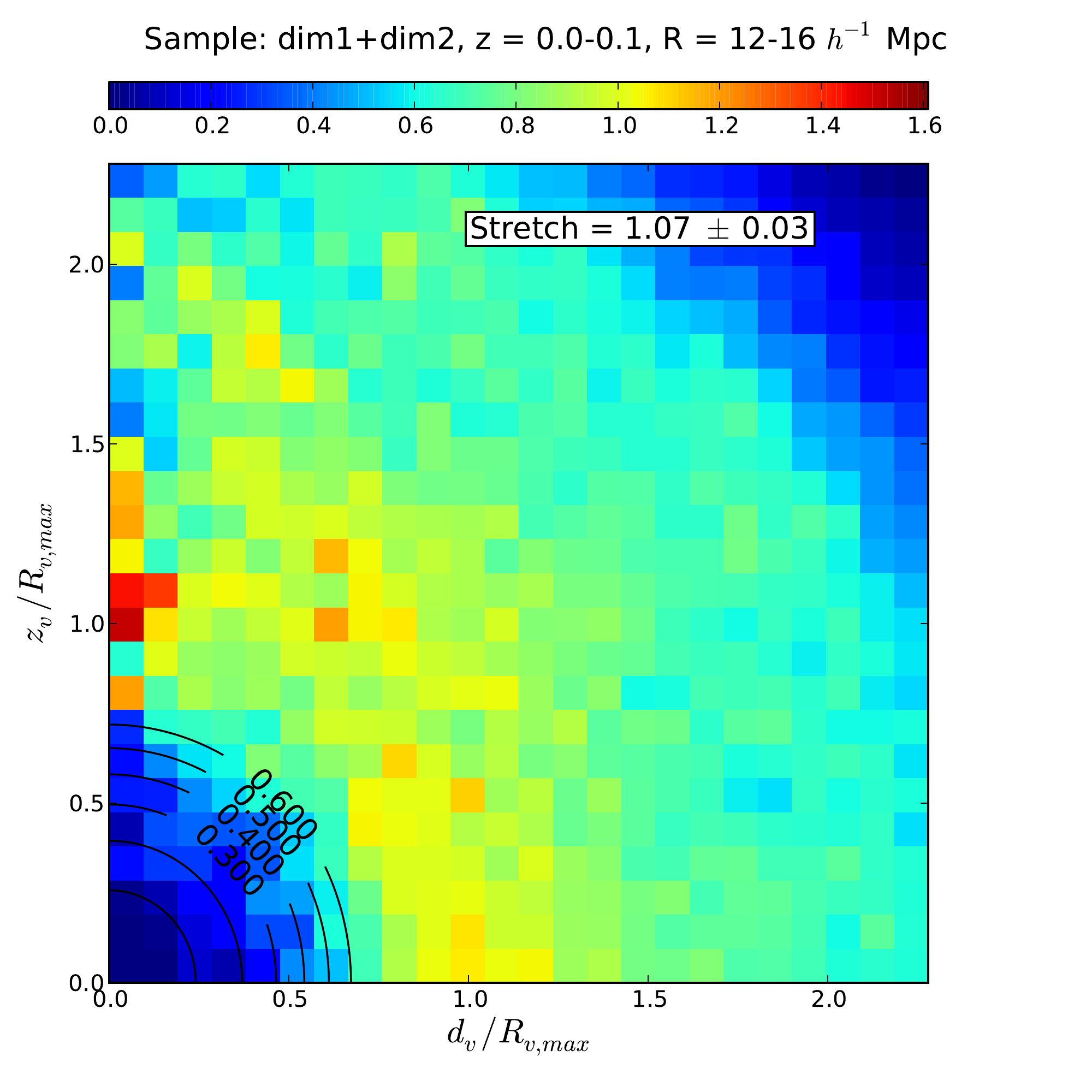}}
  {\includegraphics[type=pdf,ext=.pdf,read=.pdf,width=0.48\textwidth]{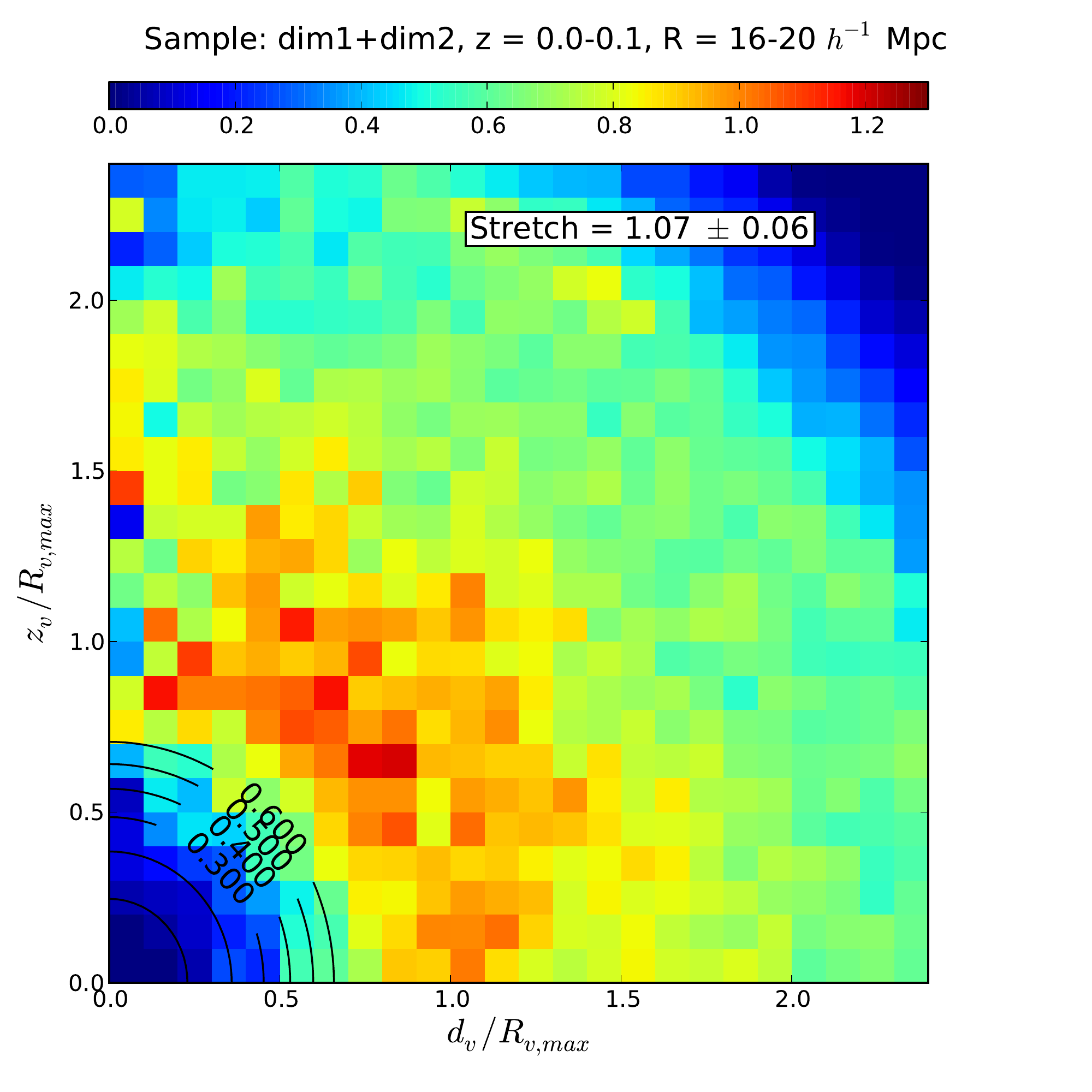}}
  {\includegraphics[type=pdf,ext=.pdf,read=.pdf,width=0.48\textwidth]{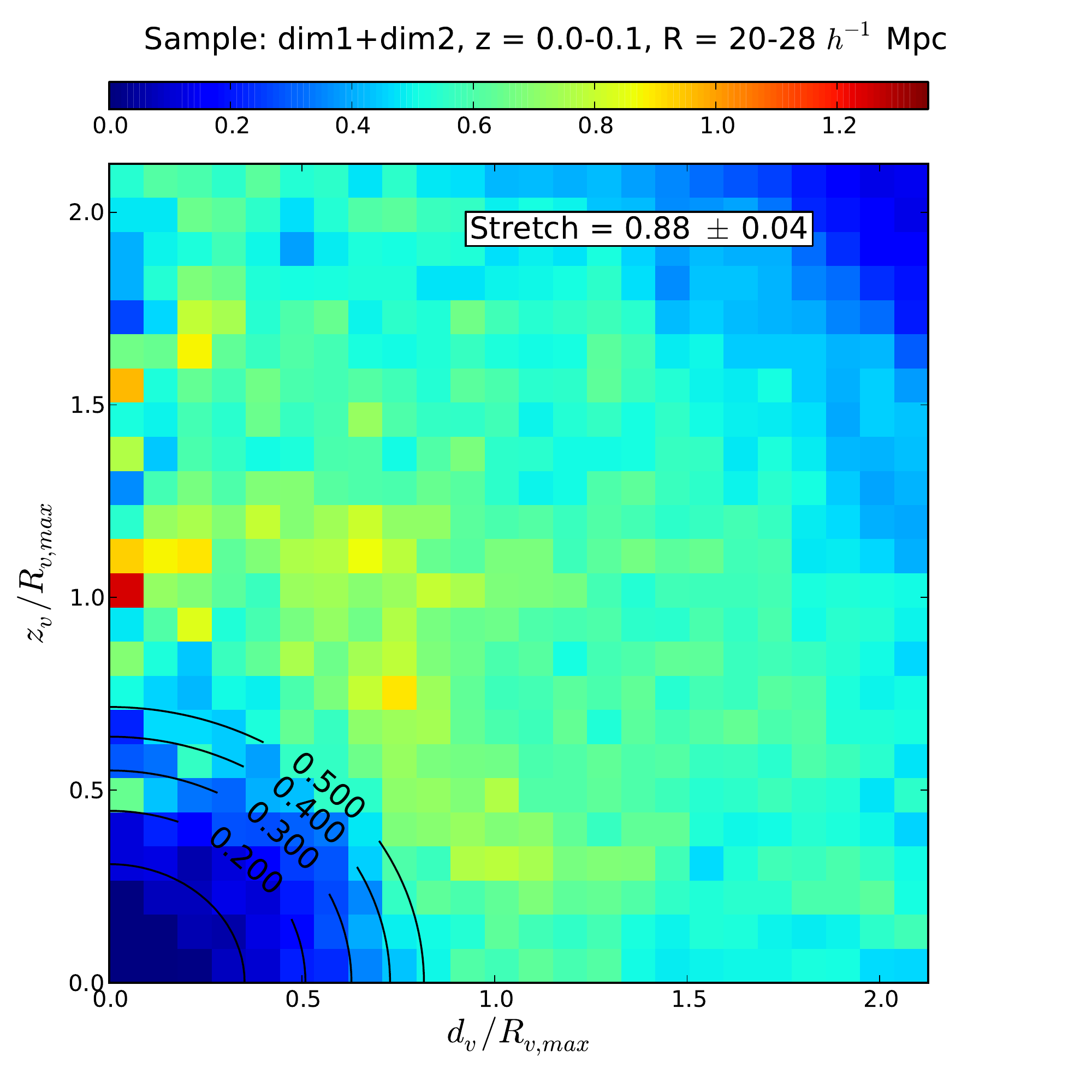}}
  \caption{\emph{Stacked voids at $0.0<z<0.1$ for the \emph{dim1+dim2} sample}.
           Shown are the stacked voids in our hybrid coordinate system
           (Equation~\ref{eq:transformation}). Colors indicate $n/\bar{n}$, 
           the number density 
           in that bin relative to the mean number density of the sample. 
           The black lines are contours of constant density of our fitted 
           ellipse for each stack, 
           and the text box indicates the measured stretch, or ratio of 
           the length along the line of sight (redshift direction) to the 
           angular extent of the fitted ellipse. The error bars quoted here
           come directly from the fitting procedure 
           (Equations~\ref{eq:ellipse}-\ref{eq:likelihood}).}
\label{fig:profile2ddim}
\end{figure*}

\begin{figure*} 
  \centering 
  {\includegraphics[type=pdf,ext=.pdf,read=.pdf,width=0.48\textwidth]{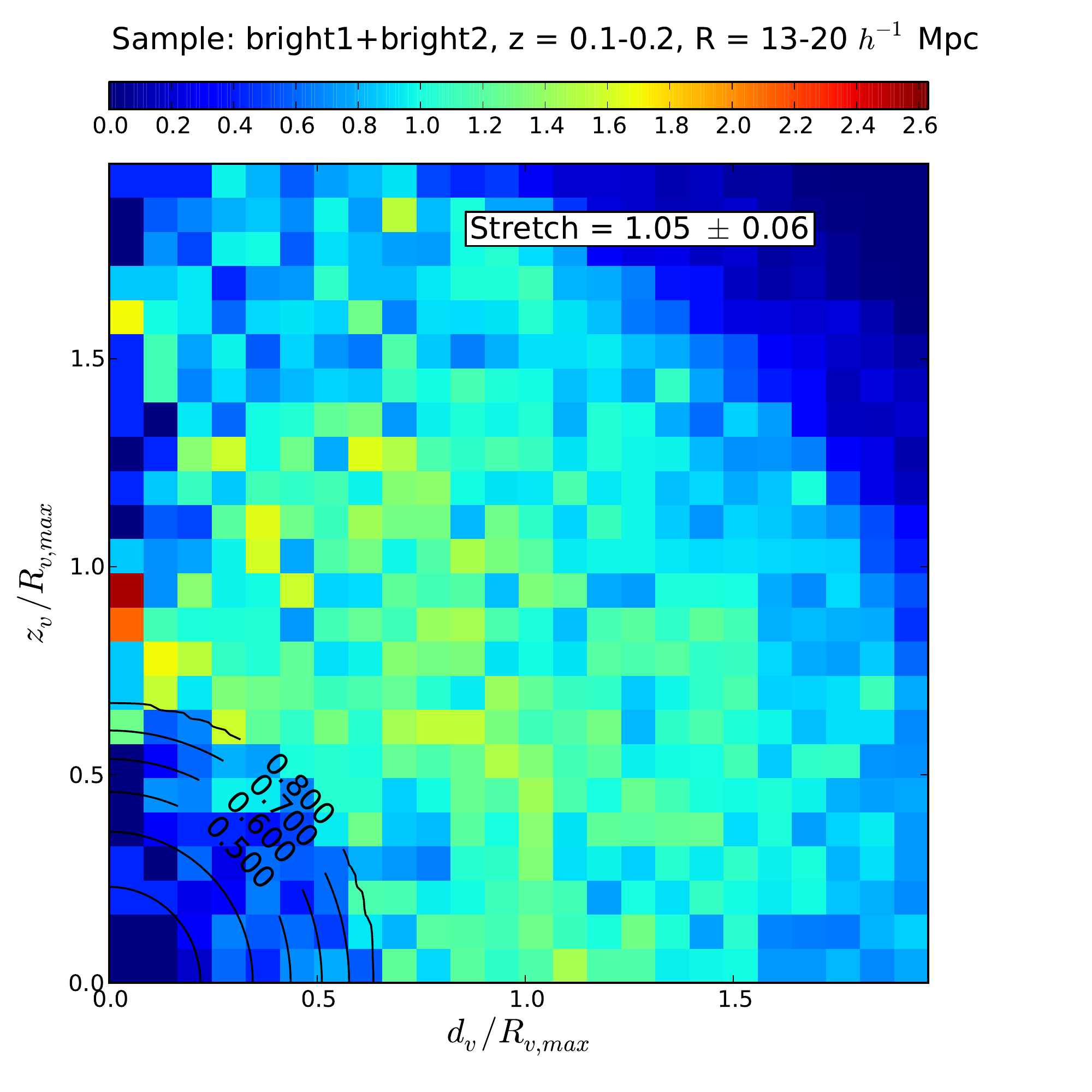}}
  {\includegraphics[type=pdf,ext=.pdf,read=.pdf,width=0.48\textwidth]{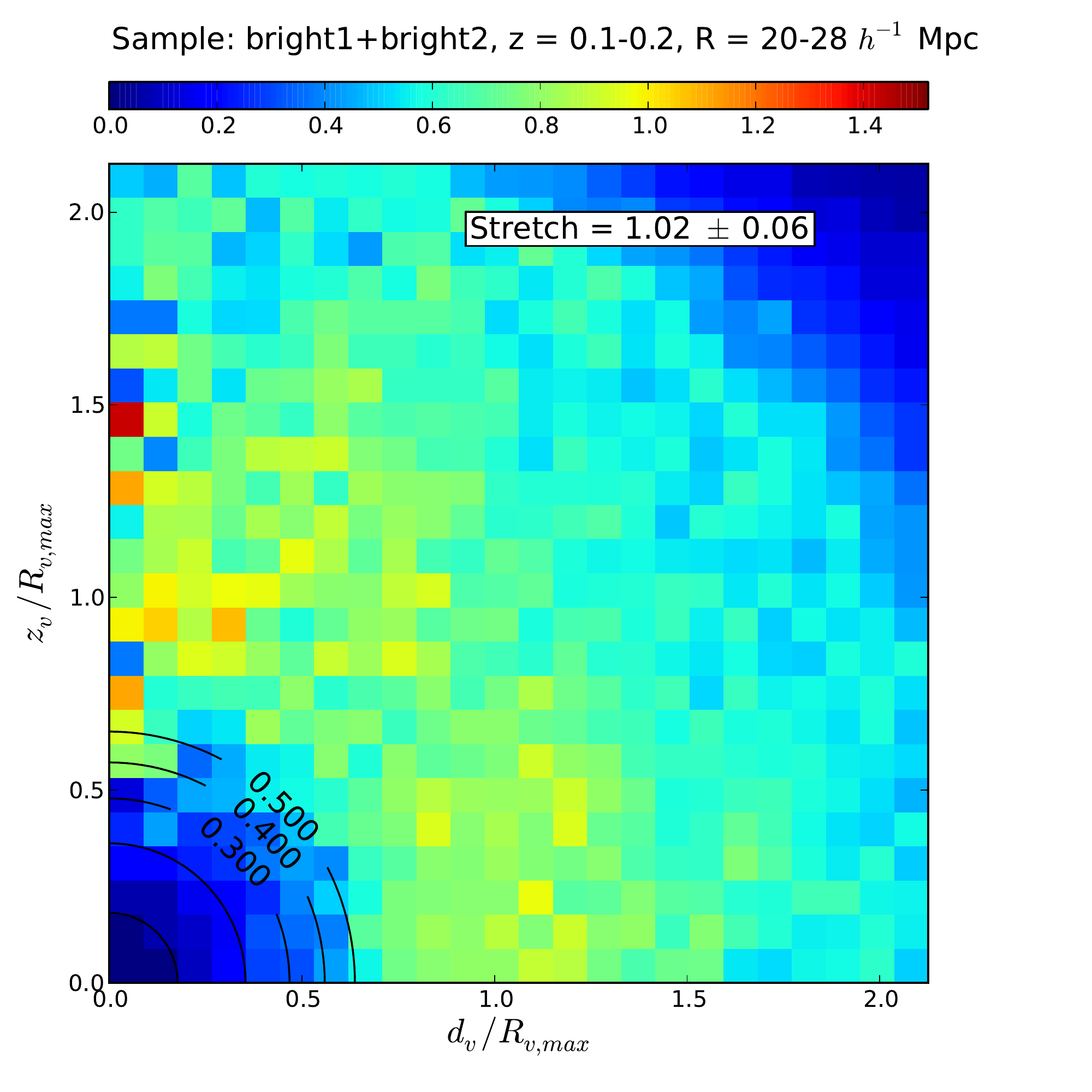}}
  {\includegraphics[type=pdf,ext=.pdf,read=.pdf,width=0.48\textwidth]{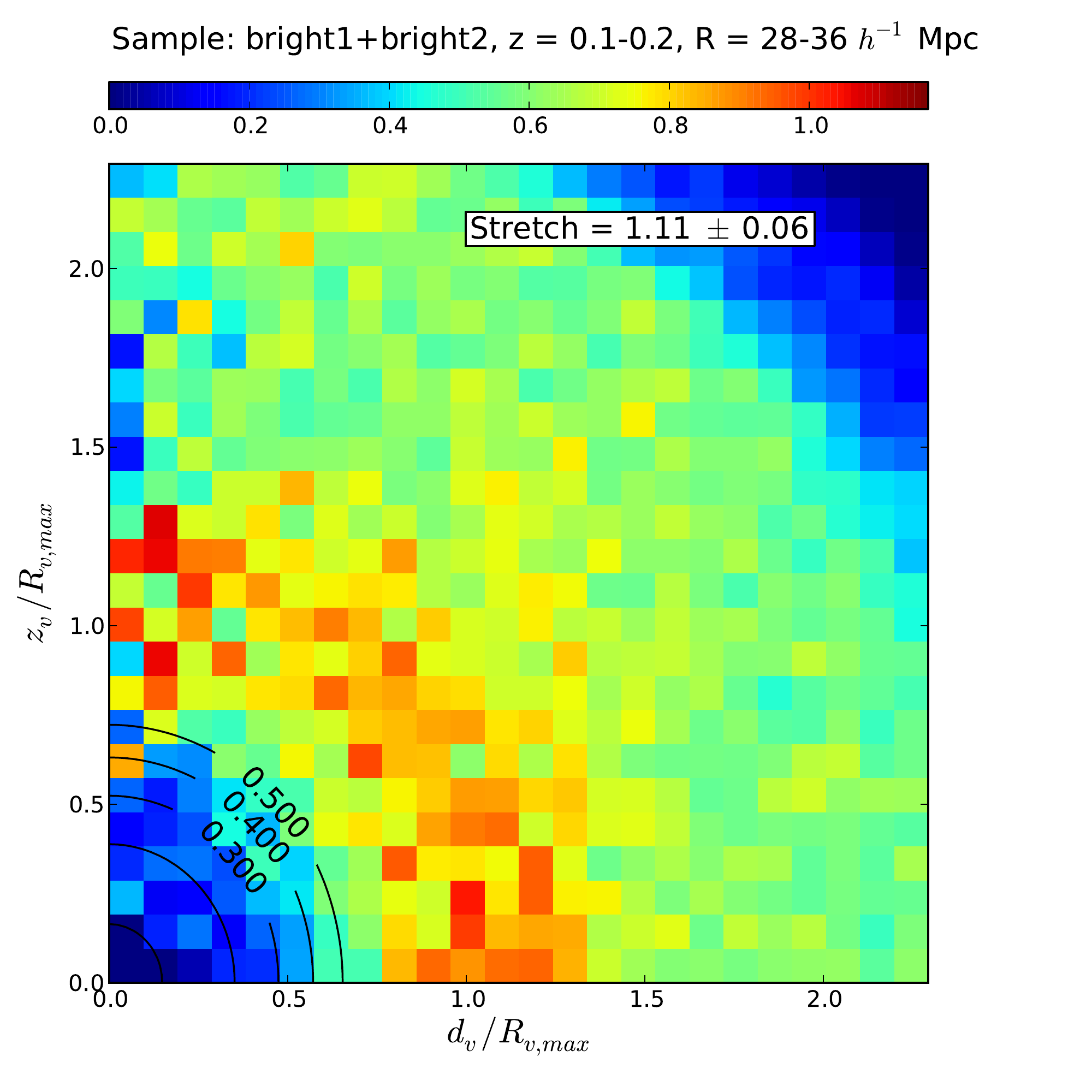}}
  {\includegraphics[type=pdf,ext=.pdf,read=.pdf,width=0.48\textwidth]{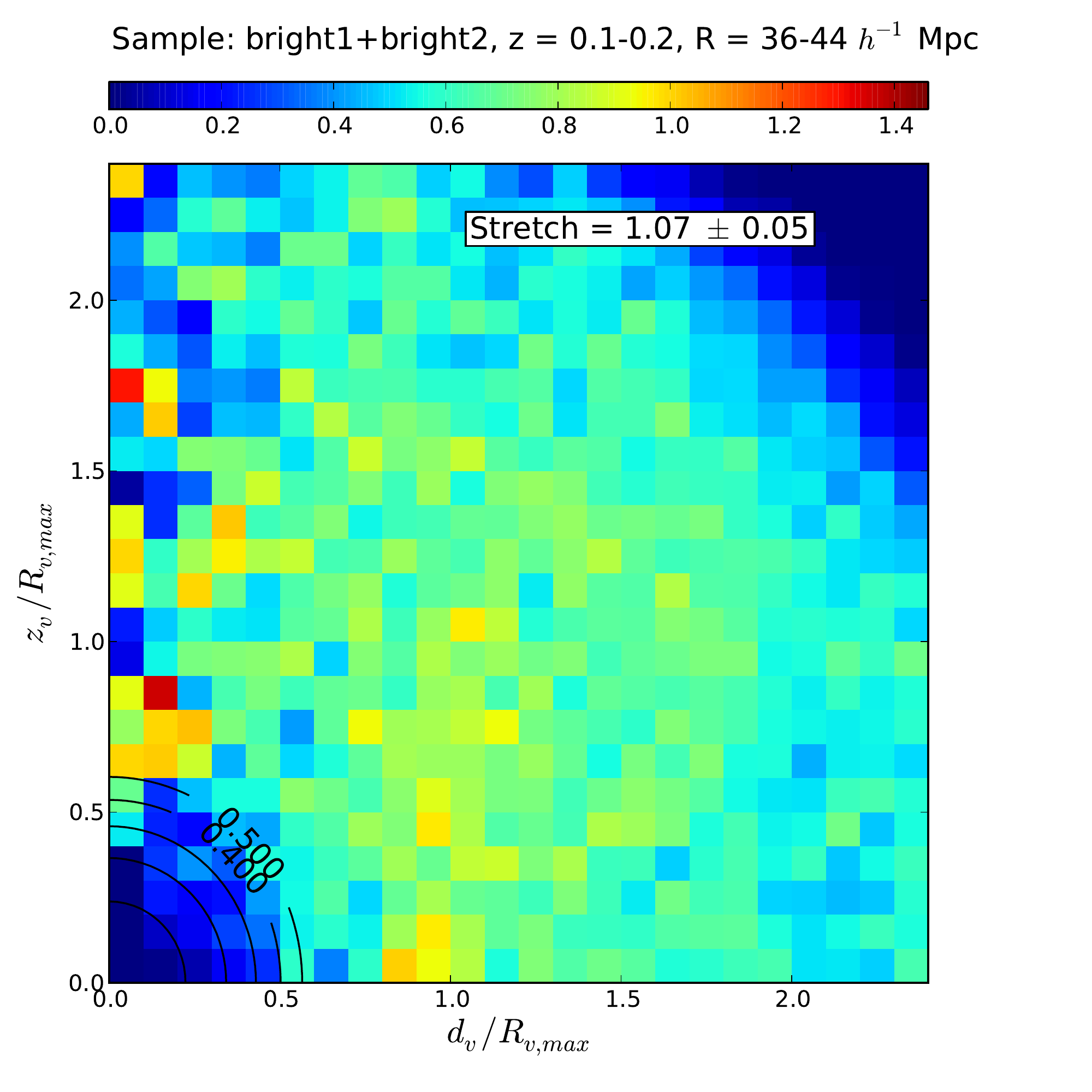}}
  \caption{\emph{Stacked voids at $0.1<z<0.2$ for the 
                 \emph{bright1+bright2} sample}.
           See the caption for Figure~\ref{fig:profile2ddim} for a plot
           description.
           }
\label{fig:profile2dbright}
\end{figure*}

\begin{figure*} 
  {\includegraphics[type=pdf,ext=.pdf,read=.pdf,width=0.48\textwidth]{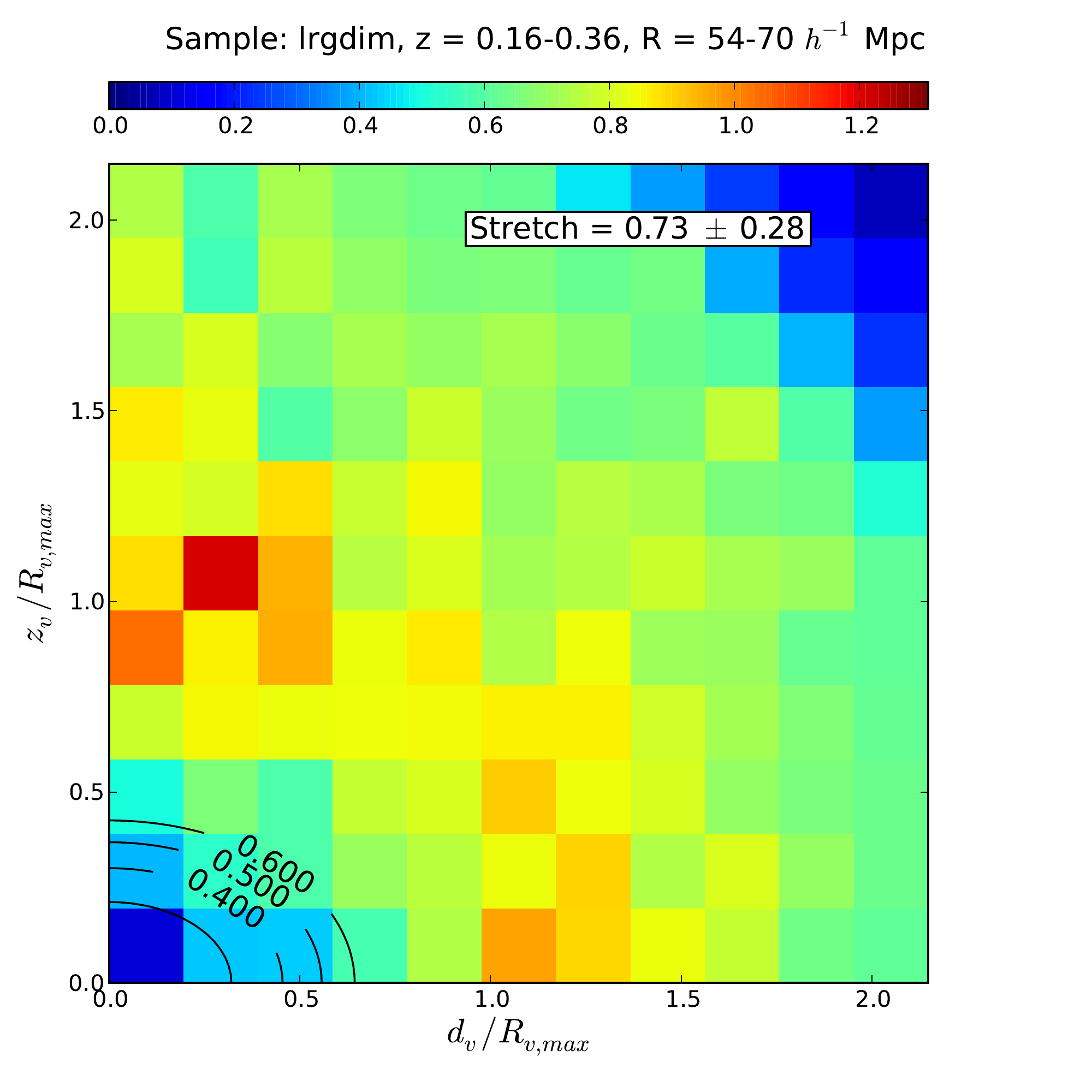}}
  {\includegraphics[type=pdf,ext=.pdf,read=.pdf,width=0.48\textwidth]{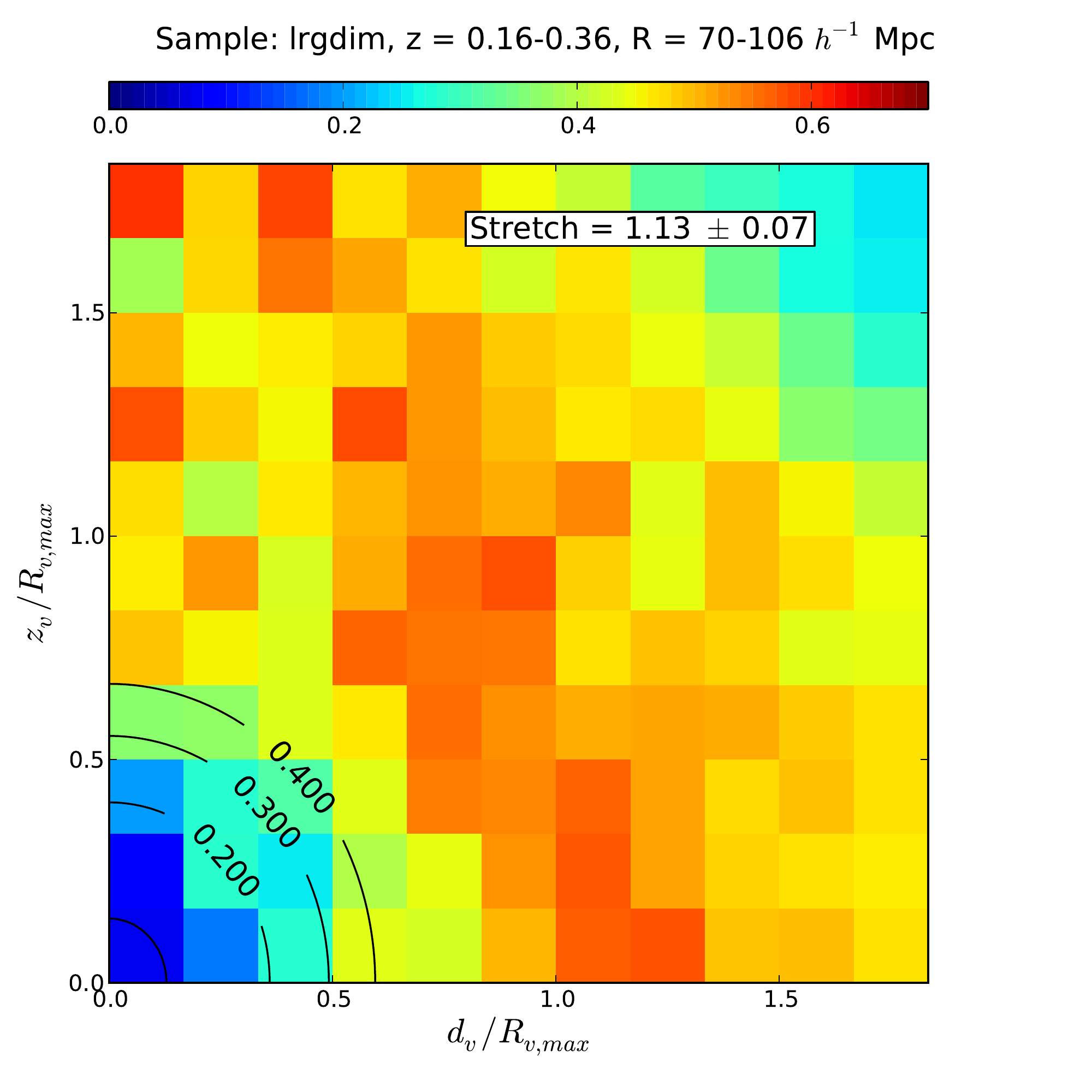}}
  \caption{\emph{Stacked voids at $0.16<z<0.36$ for the \emph{lrgdim} sample}.
           See the caption for Figure~\ref{fig:profile2ddim} for a plot
           description.
           }
\label{fig:profile2dlrgdim}
\end{figure*}

We show the two-dimensional stacks used in our shape-fitting analysis
in Figures~\ref{fig:profile2ddim},~\ref{fig:profile2dbright}, 
and~\ref{fig:profile2dlrgdim}.
As in~\citet{LavauxGuilhem2011}, to improve the signal-to-noise we fold 
the stacked void about the $d$-axis 
and to alleviate the effects of Poisson sampling we discretize the density.
While~\citet{LavauxGuilhem2011} used fixed 2~\hmpc~bins for this 
discretization, our void sizes are much larger, so we choose to 
scale the bin size with the void radius so that there are always 
10 bins within the maximum void radius.
For the \emph{lrgdim} sample we widen the pixel size to dampen the 
larger Poisson noise.
The black lines in each plot are the isodensity contours of our best-fit 
ellipse for each stack using the technique outlined above.
The contours stop at the calculated value of $n_{max}$, which is the 
mean density outside the void wall.
We also list the calculated ratio $\delta z_v/\delta d_v$ in the plot.

Even with our discretization there is still significant variation in the
density, especially outside the inner walls of the voids. Additionally, the
high-density ridge surrounding the void stack known as the ``compensation
region'' is not clearly defined, since our rescaling of individual voids is
designed to clear out the inner regions, which correspondingly widens the
compensation region. Fortunately, this does not significantly affect our
shape-fitting procedure: by assuming a steep profile, we are most sensitive to
the more clearly-defined inner edge of the wall.

Poisson noise and the small number of detected voids make 
finding a reliable fit 
in the \emph{lrgdim} samples difficult; we will see that 
this is reflected in the 
larger error bars compared to the other samples.

\section{Void stretch}
\label{sec:ellipticity}

Although the fitting procedure defined by
Equations~(\ref{eq:radial})-(\ref{eq:likelihood}) provides an 
estimate of the uncertainty in each measurement, the 
method makes rather simple assumptions about the errors in density,
namely that they are Gaussian distributed (justifying $\chi^2$
likelihood) with the errors given by Equation~(\ref{eq:stdev}).
This procedure appears to give reasonable statistical errors
in the dark matter simulations analyzed by~\citep{LavauxGuilhem2011},
but here we are analyzing galaxy catalogs, which are sparser
and to some degree biased tracers of structure.
We have therefore developed an empirical method of estimating
errors by creating ``incoherent'' void stacks:
instead of aligning voids within a stack 
to have a common line of sight, 
we assign a random set of Euler angles.
After rotating each void in this fashion, we
align the barycenters and stack the voids as usual. 
Since we have removed any information about the line of sight
by construction, the mean stretch of the coherent stack must be unity,
and the scatter about this value arises from statistical fluctuations
that include both the intrinsic scatter in void shapes and the
errors in the shape measurements.  We create an
ensemble of 100 such incoherent stacks, using the same voids in each stack but 
seeding each new stack with a unique random number seed,
and take the rms dispersion of the stretch values as our estimate
of the $1\sigma$ error.

\begin{figure} 
  \centering 
  {\includegraphics[type=pdf,ext=.pdf,read=.pdf,width=\columnwidth]
    {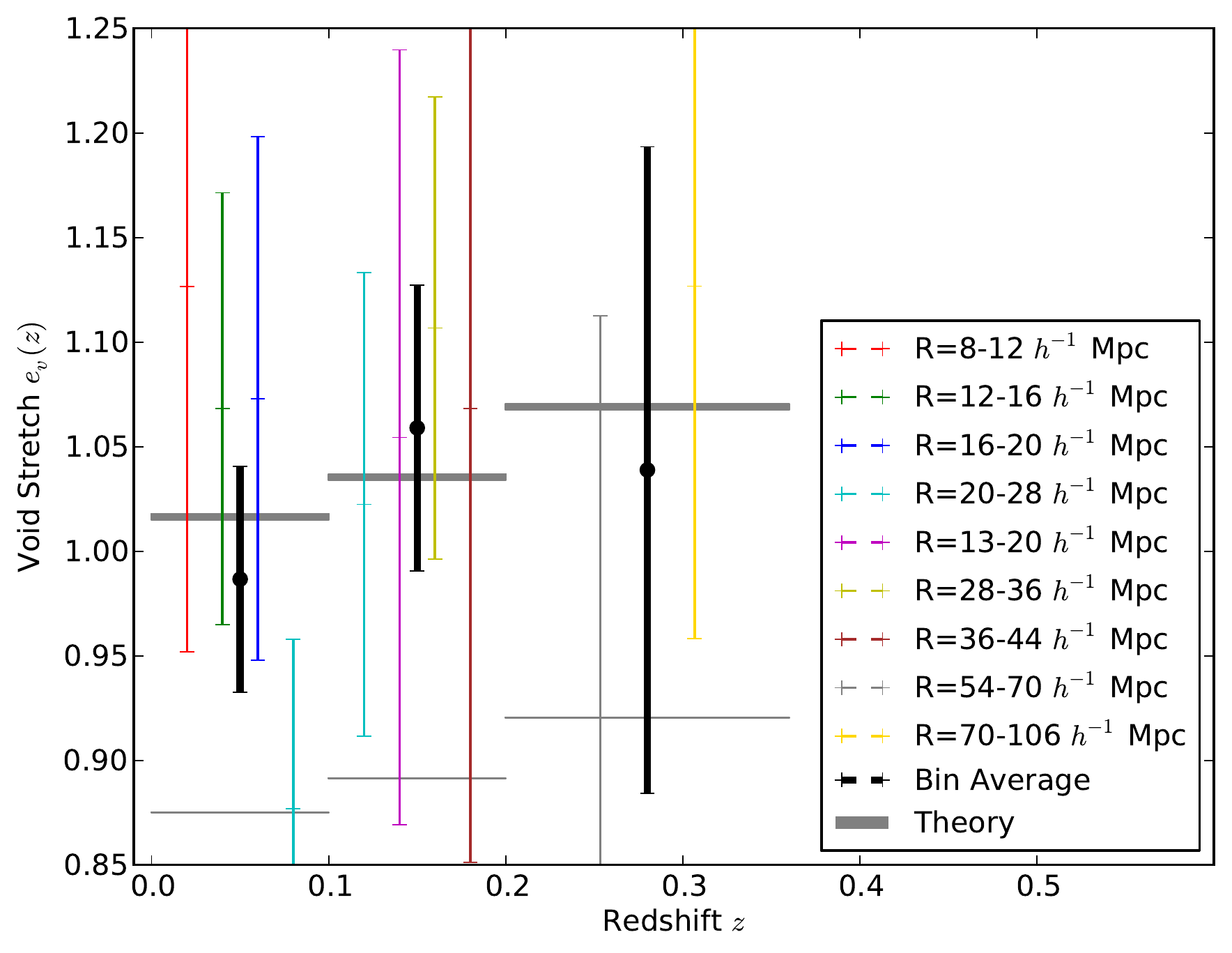}}
  \caption{\emph{Void stretch as a function of redshift.}
           We show the measured void stretch (points with error bars) 
           of each stack for all samples 
           versus the expected mean stretch in that redshift 
           bin (thick gray horizontal bands) assuming a cosmology consistent 
           with other recent observations~\citep{Komatsu2011}.
           The thin gray lines indicate the expected stretch including 
           the systematic offset of $16 \%$ induced by peculiar velocities
		   that was found in simulations by 
           \citet{LavauxGuilhem2011}.
           We assign each 
           radial bin a unique color. Also, the radial bins are ordered 
           left-to-right within each sample redshift range. 
           Note that we distribute the individual points within 
           the redshift range for clarity of plotting only.
           The black points with error bars indicate the weighted 
           mean of the measurements in that redshift range.
           Error bars indicate $1\sigma$ uncertainty and are derived from 
           an ensemble of incoherent stacks.
           Note that for clarity we have truncated the mean 
           ellipticity line of the \emph{lrgdim} sample so that it does 
           not overlap the \emph{bright1+bright2} range.
           }
\label{fig:hubble}
\end{figure}

We may construct diagrams of the measured versus expected void stretch via
the identity in Equation~(\ref{eq:ap}). 
Figure~\ref{fig:hubble} shows such a diagram where we collect our 
stretch measurements from each stack of each sample and 
compare those to the expected mean stretch in that 
redshift bin as a function of 
redshift, $\overline{e_v}(z)$ (Equation~\ref{eq:avestretch}).
The error bars shown represent the $1\sigma$ scatter in the  
100 incoherent stacks. 
For the expected stretch we assume a fiducial cosmology 
of $\Omega_M=0.27$, $\Omega_\Lambda=0.73$, and $h_0=0.71$, 
consistent with the latest WMAP 7-year results combined with 
supernovae and BAO observations~\citep{Komatsu2011}.

There is significant scatter in the \emph{dim1+dim2} sample; 
this is most likely due to the uneven distribution of voids within the 
redshift range.
We also see significant scatter in the \emph{lrgdim} sample due to the 
large amount of noise present in the stacked voids.
Unfortunately, the statistical errors are too large to detect the 
expected signal of the AP effect, an increase in stretch from
1.02 in the lowest redshift bin to 1.07 in the highest redshift
bin (thick gray bars).

Peculiar velocities have a small but not negligible effect
on average void shapes.  In N-body tests,~\citet{LavauxGuilhem2011}
find a mean bias of 1.16, with peculiar velocities systematically
flattening voids along the line of sight and reducing the
stretch factor. Applying this simple factor was all that was necessary 
to correct for systematics in their analysis and produce results consistent 
with expectations. Thin gray lines in Figure~\ref{fig:hubble}
show predictions that include this suppression, which actually
strongly disagree with our SDSS measurements.
However, \citet{LavauxGuilhem2011} considered dark matter rather than sparse,
biased galaxy tracers, and the voids in their analysis were
mostly smaller than the ones in our sample (though somewhat overlapping
in size). While correcting for systematics may indeed in the end be 
``simple'', in terms of only requiring additive or multiplicative factors 
for a
given population of voids,
the corrections may be a function of sampling density and void size.
 Further theoretical work is needed to predict
peculiar velocity effects in the regime studied here and evaluate
the disagreements with our measurements.

We also performed the same analysis as
above but including \emph{all} available voids, including truncated 
voids near the surveys edges and masks. This also includes all voids 
which, if rotated, would intersect any edges. The survey boundaries 
preferentially select voids that lie parallel to them: thus 
the mask edges will bias our results with an excess of voids 
parallel to the line of sight, while the redshift boundaries will 
bias our results with an excess of voids perpendicular to the 
line of sight. 
We found that we do not recover a strong positive AP signal, and instead 
find measurements that scatter around $\sim 0.95$.
Since the surface area of the spherical cap which defines the redshift 
boundary is two to three times greater than the surface area of the 
cone which defines the mask edges, we expect our results to be biased 
below unity when including all voids. 
We also increase the scatter in individual measurements due to the 
inclusion of many smaller and less well-resolved voids. However, 
with $\sim 30\%$ more voids in each stack we do reduce the error bars 
for each measurement by a factor of roughly $1/\sqrt{N_v}$.

\begin{figure} 
  \centering 
  {\includegraphics[type=pdf,ext=.pdf,read=.pdf,width=\columnwidth]
    {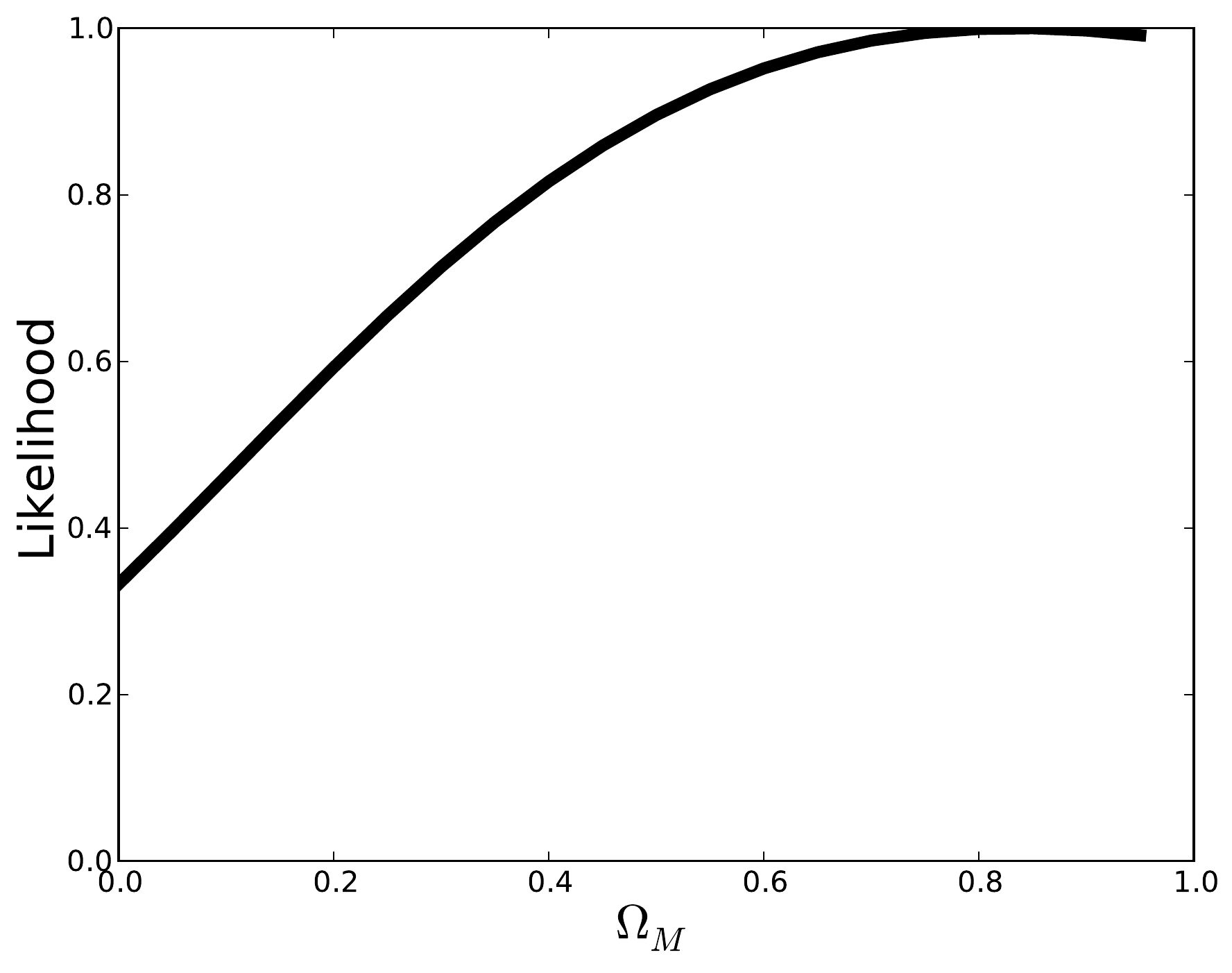}}
  \caption{\emph{One-dimensional likelihood derived from voids.}
		   We plot the relative likelihood for values of 
           $\Omega_{\rm M}$ in a flat-$\Lambda$ universe, after 
		   marginalizing over a constant peculiar velocity distortion
		   factor in the range $1-1.2$, assumed to be independent
		   of redshift.}
\label{fig:1dlikelihood}
\end{figure}

Figure~\ref{fig:1dlikelihood} shows the relative likelihood
of $\Omega_{\rm M}$ values in a flat universe with a cosmological
constant, given our stretch measurements of SDSS voids.
We calculate this likelihood using the weighted average 
measurements in each redshift bin (black points in Figure~\ref{fig:hubble}),
assuming a Gaussian likelihood function.  To allow for the effects
of peculiar velocities, we marginalize over a constant multiplicative
bias factor with a uniform prior in the range $[1.0-1.2]$
(the same range of values assumed in the forecasts
of ~\citealt{LavauxGuilhem2011}). 
A positive detection of the AP effect would correspond to a rejection
of $\Omega_{\rm M}=0$, since a flat, pure-$\Lambda$ universe has
constant $H(z)=H_0$ and $D_A(z)=cz / H_0$
(i.e., it really does have the coordinate system of 
Equation~\ref{eq:transform}).
As expected from Figure~\ref{fig:hubble}, our current statistical
errors are too large compared to the predicted stretch signal
to detect the AP effect.

\section{Conclusions}
\label{sec:conclusions}

We have performed the first application of the  
Alcock-Paczynski test to stacked voids to observational data. 
We applied the AP test by 
measuring ellipticities of stacked voids using the
void catalog of~\citet{Sutter2012}. 
The stacking procedure greatly reduces the effects of Poisson noise 
and allows us to reliably apply the shape-fitting algorithm 
of~\citet{LavauxGuilhem2011}.
By grouping voids into multiple 
radial bins we obtain many independent measurements, and by dividing the 
void catalog into redshift bins we obtain measurements across the full 
range of the SDSS DR7 main sample and most of the LRG catalog.
However, the limited number of voids and the considerable
scatter that remains does not allow us to positively identify
the AP effect over the redshift range probed by
these data.

The SDSS-III BOSS survey~\citep{Dawson2012} should be a much more
powerful basis for void-based AP measurements than the DR7 redshift
survey analyzed here.  First, in the range of redshift overlap with
our {\it lrgdim} sample ($z=0.16-0.36$), the space density of
BOSS galaxies is a factor $\sim 3$ higher, which enables identification
of smaller (and more numerous) voids and more accurate measurement
of void density distributions.  Second, BOSS extends this higher
sampling density out to $z\approx 0.65$, probing a larger comoving
volume (and hence more voids) and reaching redshifts where the predicted
AP signal is larger, with a stretch factor $e_V(z=0.65) = 1.2$ for
a flat-$\Lambda$ universe with $\Omega_{\rm M}=0.27$.  Future
ground-based surveys like BigBOSS~\citep{Schlegel2011} could extend
these studies to $z\approx 1$, while the space-based emission
line redshift surveys of Euclid~\citep{Laureijs2011} and
WFIRST \citep{Green2011} will probe much larger comoving volumes
at $z=1-2.5$.  The Fisher matrix analysis of~\citet{LavauxGuilhem2011}
implies that a void-AP analysis of the Euclid survey should yield
significantly tighter dark energy constraints than the BAO 
analysis of the same survey, with a factor of ten improvement
in the Dark Energy Task Force~\citep{AlbrechtAndreas2006} Figure of Merit.

Our initial foray into observational application of this approach
highlights two important directions for future investigation.
The first is a more detailed study of measurement and parameter
fitting procedures and error estimation techniques.
Our fitting methods are closely modeled on those of~\citet{LavauxGuilhem2011},
but there may be other approaches that make better use of the
available information, such as using an empirical radial profile
in place of our adopted parametric model, changing the radial range of
the fit, or downweighting the fluctuations arising from clustered
galaxies at the void boundaries.  Alternatively, one could avoid
profile fitting entirely and instead use anisotropy of the
``void-galaxy cross-correlation function,'' analogous to the
cluster-galaxy cross-correlation but centered at density minima
instead of density maxima.  The second direction is a more detailed
study of peculiar velocity effects on mean void shapes, examining its
dependence on void size and on the spatial and velocity bias of
galaxy tracers.  The likelihood analysis in Figure~\ref{fig:1dlikelihood}
allows for an overall velocity distortion factor and therefore
effectively uses just the redshift dependence of the signal in
Figure~\ref{fig:hubble} to constrain cosmology, which was appropriate given 
the systematic effects noticed in~\citep{LavauxGuilhem2011}. 
The goal for
future analyses should be to apply a theoretically computed
velocity distortion correction to each void sample in each redshift
bin and marginalize only over the uncertainty in this correction,
getting an absolute constraint on the average void stretch, and 
hence $H(z)D_A(z)$, at each redshift.
In the context of halo occupation distribution (HOD) models
(e.g.,~\citealt{Zehavi2011}), we expect galaxies to have the 
same mean velocities as their parent halos on average, but the velocity
dispersion of galaxies could differ from that of the dark matter.
This velocity dispersion bias can itself be constrained by 
redshift-space galaxy clustering~\citep{Tinker2006}, so we expect
the residual uncertainty in peculiar velocity corrections to void
shapes to be small, though it may still be the limiting systematic
in void-based AP analysis.

The statistical errors of this approach are limited only by the
size and redshift range of spectroscopic galaxy surveys, which
are expected to grow dramatically in the coming years.
Cosmic voids are the converse of galaxy clusters; primordial
density minima expand and deepen to form non-linear structures
that fill much of the universe and are, in a sense, the most
``dark energy dominated'' regions of the cosmos.
The mean shapes of these regions may ultimately provide
powerful clues to the nature of the dark energy that pervades them.

\section*{Acknowledgments}

PMS and BDW acknowledge support from NSF Grant AST-0908902.
GL acknowledges support from CITA National Fellowship and financial
support from the Government of Canada Post-Doctoral Research Fellowship.
Research at Perimeter Institute is supported by the Government of Canada
through Industry Canada
 and by the Province of Ontario through the Ministry of Research and
Innovation.
DW acknowledges support from NSF Grant AST-1009505 and the 
hospitality of the Institut d'Astrophysique de Paris.
This material is based upon work supported in part by NSF Grant 
AST-1066293 and the hospitality of the Aspen Center for Physics.

Funding for the Sloan Digital Sky Survey (SDSS) was provided by the Alfred P. Sloan Foundation, the Participating Institutions, the National Aeronautics and Space Administration, the National Science Foundation, the U.S. Department of Energy, the Japanese Monbukagakusho, and the Max Planck Society. The SDSS Web site is http://www.sdss.org/.
\bibliography{sdssaptest}		
\bibliographystyle{apj}	\nocite{*}

\end{document}